\documentclass{article}

\PassOptionsToPackage{numbers, compress}{natbib}
 \usepackage[preprint]{neurips_2026}

\usepackage[utf8]{inputenc} 
\usepackage[T1]{fontenc}    
\usepackage{hyperref}       
\usepackage{url}            
\usepackage{booktabs}       
\usepackage{amsfonts}       
\usepackage{nicefrac}       
\usepackage{microtype}      
\usepackage{xcolor}         
\usepackage{amsmath}
\usepackage{amssymb}
\usepackage{multirow}
\usepackage{arydshln}
\usepackage{graphicx}
\usepackage{courier}
\definecolor{losscolor}{RGB}{72,112,150}    
\definecolor{gaincolor}{RGB}{165,95,80}     
\usepackage{wrapfig}
\usepackage{tcolorbox}
\tcbuselibrary{listings,breakable,skins}
\usepackage{listings}

\definecolor{boxgraybg}{RGB}{248,248,248}
\definecolor{boxgrayframe}{RGB}{210,210,210}
\definecolor{myred}{RGB}{180,35,35}

\newcommand{\gain}[1]{\hspace{0.05em}{\tiny\textcolor{gaincolor}{(#1)}}}
\newcommand{\loss}[1]{\hspace{0.05em}{\tiny\textcolor{losscolor}{(#1)}}}
\usepackage{array}
\newcolumntype{C}[1]{>{\centering\arraybackslash}p{#1}}
\newcolumntype{L}[1]{>{\raggedright\arraybackslash}p{#1}}
\usepackage[table]{xcolor}
\definecolor{groupgray}{gray}{0.90}
\usepackage{pifont}
\usepackage{float}
\usepackage{threeparttable}

\usepackage{longtable}
\usepackage[table]{xcolor}

\newcolumntype{P}[1]{>{\raggedright\arraybackslash}p{#1}}

\newcommand{\sqlfont}{\ttfamily}

\newcommand{\field}[1]{\textbf{#1}}

\definecolor{sqlbg}{RGB}{245,245,245}
\definecolor{sqlhlbg}{RGB}{255,236,236}
\definecolor{trajok}{RGB}{24,114,62}
\definecolor{trajwarn}{RGB}{153,102,0}
\definecolor{trajbad}{RGB}{160,32,32}

\newlength{\sqlhlwidth}

\newcommand{\sqlblock}[1]{%
    \begingroup
    \setlength{\fboxsep}{6pt}%
    \par\smallskip\noindent
    \colorbox{sqlbg}{\parbox{\dimexpr\linewidth-2\fboxsep\relax}{\sqlfont\scriptsize #1}}%
    \par\smallskip
    \endgroup
}

\newcommand{\sqlline}[1]{\detokenize{#1}\\}

\newcommand{\sqlcolorline}[1]{%
    \begingroup
    \setlength{\fboxsep}{1pt}%
    \colorbox{sqlhlbg}{\sqlfont\scriptsize\strut\detokenize{#1}}\\
    \endgroup
}

\newcommand{\sqlhighlight}[1]{%
    \begingroup
    \settowidth{\sqlhlwidth}{\sqlfont\scriptsize\detokenize{#1}}%
    \rlap{\raisebox{-0.45ex}[0pt][0pt]{\textcolor{sqlhlbg}{\rule{\sqlhlwidth}{2.35ex}}}}%
    \sqlfont\scriptsize\detokenize{#1}%
    \endgroup
}

\newcommand{\sqlredline}[1]{\sqlhighlight{#1}\\}

\title{Diff-SQL: SQL Efficiency Optimization via Patch Generation and Constraint Alignment}

\author{%
\textbf{Shipei Lin$^{1}$, Duomin Zhang$^{1}$, Xiaolong Li$^{2}$, Bohan Hu$^{1}$,} \\
\textbf{Bowen Qin$^{3}$, Jinyang Li$^{2}$, Chenhao Ma$^{1}$\thanks{Corresponding author: Chenhao Ma.}
} \\[1mm]
$^{1}$The Chinese University of Hong Kong, Shenzhen \\
$^{2}$The University of Hong Kong \quad
$^{3}$National University of Singapore \\
\texttt{shipeilin@link.cuhk.edu.cn}\\
\url{https://github.com/L-nshpp/Diff-SQL}
}

\begin{document}

\maketitle

\begin{abstract}
SQL efficiency optimization aims to transform slow queries into semantically equivalent but faster alternatives. However, directly optimizing SQL with large language models in an end-to-end fashion often induces \textbf{\textit{Objective Misalignment}} which creates a fundamental tension between optimization and correctness, making direct full SQL rewriting unreliable for execution-facing database applications. To address this problem, we propose \textbf{Diff-SQL}, a two-stage framework that decouples efficiency-oriented optimization from constraint-aware alignment. The first stage identifies optimization opportunities and proposes targeted edits in the form of a unified diff patch, while the second stage is trained with on-policy reinforcement learning to revise outputs under executability and semantic-equivalence constraints. To train and evaluate Diff-SQL, we construct an automated pipeline that mines optimization knowledge from StackOverflow and builds \textbf{Slow-Fast SQL pairs} through cascaded filtering. We further introduce \textbf{Effi-SQL}, a benchmark containing 1,100 human-verified Slow-Fast pairs across five SQL dialects. Experiments show that \textit{Objective Misalignment} is widespread across existing LLM-based SQL optimization methods, where direct full SQL optimization causes an average 22.7\% execution accuracy degradation across frontier models such as Claude-Opus-4.6, with the worst model dropping by 43.0\%. Diff-SQL alleviates this trade-off. As an inference-only strategy, it improves R-VES by 10.0\% on average while reducing execution accuracy degradation by 6.11\% on average across three strong base models. With execution-grounded training, Diff-SQL further enables a 7B model to improve R-VES from 33.42\% to 46.83\%, demonstrating that the proposed two-stage optimization-and-alignment paradigm can deliver both stronger efficiency and better correctness in local, small model deployment settings.
\end{abstract}

\section{Introduction}
\label{sec:introduction}
In modern database systems, SQL optimization is a fundamental lever for performance tuning, moving beyond mere syntactic validity to the pursuit of execution efficiency. While traditional query optimizers employ rule-based heuristics designed by human experts to transform queries, they often struggle with the sprawling complexity of real-world production workloads and diverse physical designs~\citep{DBLP:journals/corr/abs-2506-07675}. In cloud-native environments like Microsoft Azure SQL or Google Cloud SQL, a suboptimal SQL structure is no longer a minor quality issue; it is a primary driver of resource exhaustion and Service-Level Agreement (SLA) violations. As the industry shifts toward autonomous database management, SQL rewriting has evolved from an intermittent manual tuning exercise into a high-stakes, execution-facing requirement. This calls for a transition from static, rule-bound rewrites to a more dynamic, environment-aware optimization paradigm that can guarantee both semantic equivalence and tangible performance gains under live execution contexts as in broader software engineering tasks~\citep{DBLP:journals/corr/abs-2507-12415,DBLP:conf/nips/0005DWWQCG024}.

\begin{figure}[t]
    \centering
    \includegraphics[draft=false,width=1.0\linewidth]{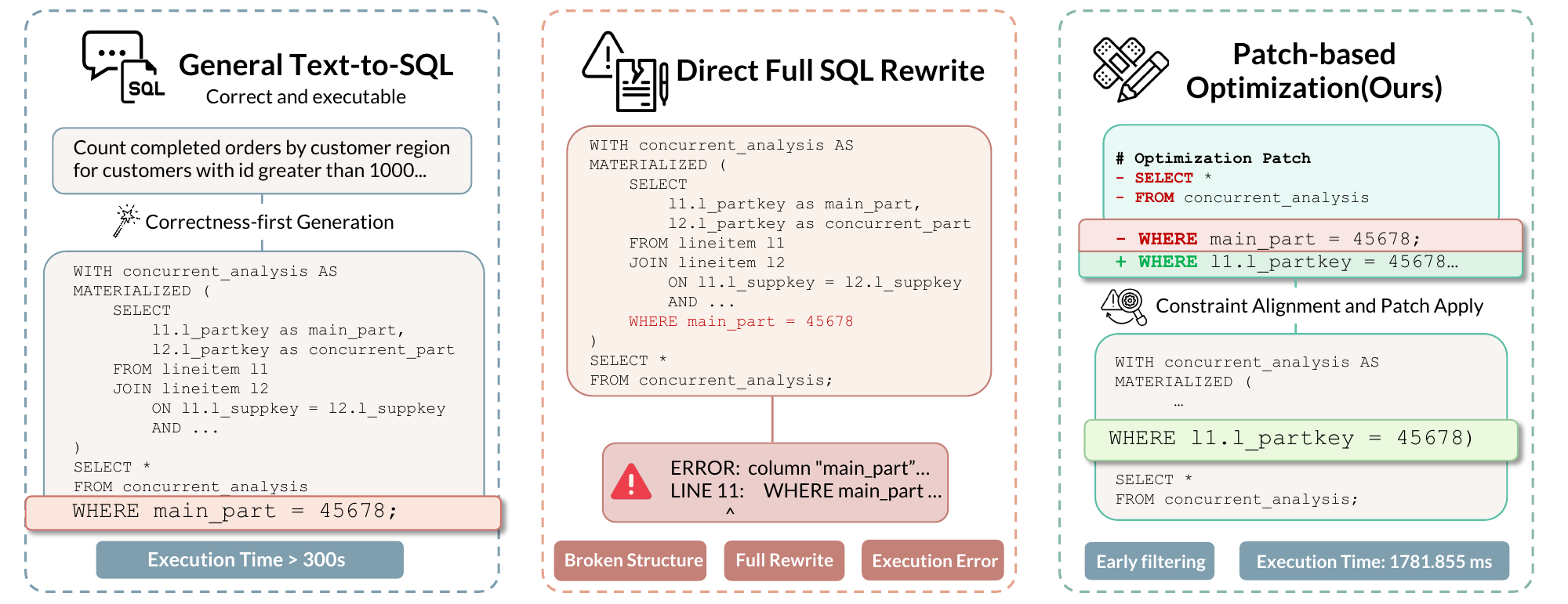}
    \vspace{-2 mm}
    \caption{Comparison between correctness-first Text-to-SQL, direct full query rewriting, and our \textbf{patch-based SQL optimization framework}. 
A correct Text-to-SQL query can still be inefficient under a concrete execution environment, while direct full query rewriting may introduce structural drift and execution errors. 
Our patch-based formulation localizes the optimization to the relevant predicate and applies constraint alignment before patch application, improving execution behavior while preserving the verified query structure.
}
    \vspace{-5 mm}
    \label{fig:task_definition}
\end{figure}

The emergence of Large Language Models (LLMs) offers a promising avenue for automating complex SQL rewrites~\citep{DBLP:journals/corr/abs-2506-07675,DBLP:conf/edbt/DharwadaDHD26,DBLP:journals/corr/abs-2508-09023} that often fall outside the fixed rewrite spaces of traditional rule-based optimizers~\citep{DBLP:journals/pvldb/SunZLYFZ25,DBLP:journals/pvldb/LiYWCB24}. Their strong performance on related tasks such as Text-to-SQL~\citep{DBLP:conf/iclr/PourrezaL0CTKGS25,DBLP:journals/tkde/LiuZGLLSHGLDZ26,DBLP:journals/corr/abs-2503-23157,DBLP:conf/acl/QuLLQH0C24,DBLP:journals/pvldb/LiWZHZJWZCSCL25,DBLP:journals/corr/abs-2408-07702,DBLP:conf/ijcnlp/ShengS25a,DBLP:conf/icml/Li0FXC0L25,DBLP:journals/corr/abs-2405-16755,DBLP:journals/corr/abs-2409-16751,DBLP:journals/corr/abs-2411-00073,DBLP:conf/coling/LeePKP25} suggests that LLMs can reason over schemas, query intent, and SQL structure. However, this capability does not directly convert into reliable SQL optimization, because producing a faster rewrite requires not only generating valid SQL, but also preserving semantics while improving execution behavior under a particular database environment, as illustrated in the left panel of Figure~\ref{fig:task_definition}. In other words, the ability to generate correct code is not sufficient for producing efficient SQL rewrites~\citep{DBLP:conf/euromlsys/EizaguirreTA26}. Unlike standard code generation, the ``goodness'' of a SQL optimization is inherently environment-dependent: it is shaped by physical schema design, data distributions, available indexes, and engine-specific execution plans. Consequently, training LLMs to optimize SQL requires a fundamental shift from learning ``how to write'' syntactically valid queries to learning ``how to perform'' within a specific execution substrate~\citep{Deochake2025CostTO}.

Learning such efficiency-aware behavior presents two fundamental challenges. First, an efficient rewrite cannot be judged by query text alone; it requires comparative, execution-grounded labels. Obtaining high-quality, verified SQL pairs at scale, where one SQL query is demonstrably faster than its semantically equivalent counterpart, is labor-intensive and error-prone. Second, as illustrated in the middle panel of Figure~\ref{fig:task_definition}, standard LLMs typically adopt an end-to-end regeneration approach. By rewriting the entire SQL from scratch to achieve localized efficiency gains, they often introduce execution errors in regions that were originally correct. This ``over-rewrite'' tendency makes LLMs difficult to trust in production environments where functional correctness is non-negotiable.

To address these challenges, we build a complete pipeline for SQL efficiency optimization. Our main contributions are as follows: (1) \textbf{Automated Scaffold of Slow-Fast SQL Pairs.} We design an automated scaffold that converts noisy community optimization heuristics into execution-verified Slow-Fast SQL pairs. (2) \textbf{Diff-SQL.} We propose a two-stage patch-based framework for semantics-preserving SQL efficiency optimization. It decouples patch generation from constraint alignment in a shared unified-diff patch space, reducing the brittleness of direct full SQL rewrite under strict executability and semantic-consistency constraints, as shown in the right panel of Figure~\ref{fig:task_definition}. (3) \textbf{Effi-SQL.} We construct a new multi-dialect benchmark for SQL efficiency optimization, containing 1,100 Slow-Fast SQL pairs. Each pair is verified through both execution and human inspection.

Our experiments on \textbf{Effi-SQL} show that SQL optimization is consistently constrained by \textbf{\textit{Objective Misalignment}} for all frontier LLMs, where models that aggressively pursue efficiency gains tend to over-commit to structural changes at the expense of semantic correctness, leading to significant execution accuracy degradation, with $\Delta$EX reaching 43.00\%. This observation motivates decomposing optimization into two stages, separating efficiency-oriented patch generation from constraint-aware refinement to alleviate this tension. In the inference-only setting, \textbf{Diff-ICL} validates this decomposition as a general prompting framework across frontier LLMs, yielding up to 12.42\% R-VES improvement while simultaneously reducing $\Delta$EX by up to 11.00\%. For deployment scenarios where data privacy requires local inference, we further train a 7B-scale model with \textbf{Diff-RL}, which improves R-VES from 33.42\% to 46.83\% and reduces $\Delta$EX from 43.67\% to 28.67\%, on par with the performance of larger frontier models on \textbf{Effi-SQL}.

\section{Preliminaries}
\paragraph{SQL optimization.}
We study SQL optimization for execution efficiency. Let $q$ be a slow SQL query and $\mathcal{X}_q = \{S_q, C_q, P_q\}$ be its corresponding context, where $S_q$, $C_q$, and $P_q$ denote the database schema, column statistics, and the QEP (query execution plan) of $q$, respectively. We aim to find an optimized SQL $q'$. The optimization mapping is defined as:
\begin{equation}
    q' = f(q, \mathcal{X}_q).
\end{equation}

To evaluate the efficiency of $q'$, we rely on real execution on the database instance $D$. The output SQL $q'$ is considered a valid optimization if it satisfies both correctness and speedup constraints:
\begin{equation}
    \quad \text{Equiv}(q, q'; D) = 1, \quad \text{and} \quad T(q'; D) < T(q; D),
\end{equation}
where $T(\cdot; D)$ denotes the execution time on $D$, $\mathrm{Equiv}(q,q';D)$ measures whether $q'$ is equivalent to $q$ on $D$.

\section{Slow-Fast pairs construction and benchmark}
\label{sec:pairs construction}
\paragraph{Overview.}
To reduce human expert labor, we introduce an automated scaffold that collects source SQL, generates diverse optimization candidates guided by distilled heuristics, and constructs Slow-Fast pairs through execution on real database engines, where each pair consists of semantically equivalent SQL queries ordered by execution latency under the same database environment.

\subsection{Automated scaffold of Slow-Fast pairs.}
\label{sec:Automated scaffold}
\begin{figure}[t]
    \centering
    \includegraphics[draft=false,width=1.0\linewidth,trim=5mm 0mm 4mm 0mm,clip]{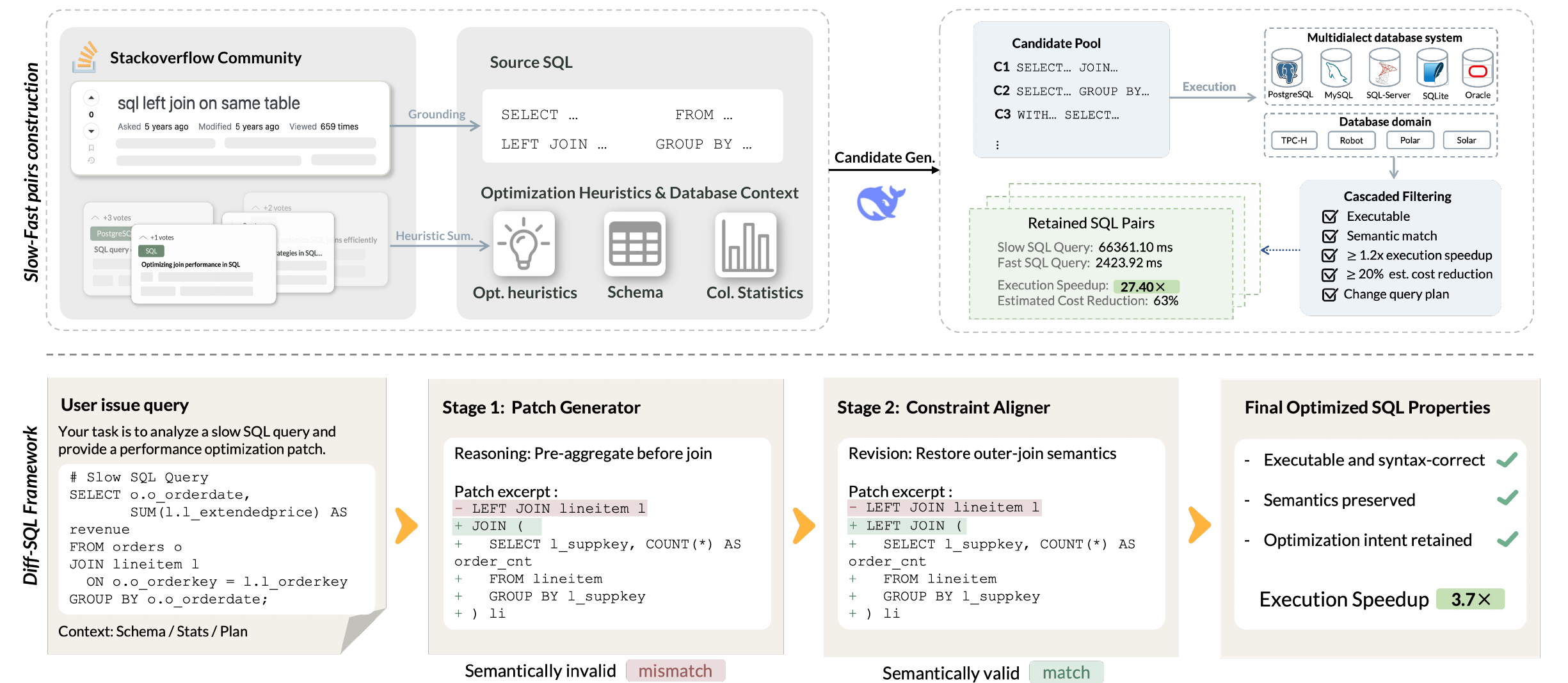}
    \vspace{-2 mm}
    \caption{Overview of \textbf{Slow-Fast pair construction} and the \textbf{Diff-SQL} framework.
The upper panel shows how we construct Slow-Fast SQL pairs by collecting performance-related SQL discussions, grounding source queries in controlled database, generating diverse optimization candidates, and retaining only pairs that pass executability, semantic-equivalence, speedup, cost-reduction, and plan-change filters.
The lower panel shows how Diff-SQL first generates an optimization reasoning and patch, then refines it into an executable and semantically correct final rewrite.}
    \vspace{-5 mm}
    \label{fig:main graph}
\end{figure}

\paragraph{Source SQL collection.}
As illustrated in Figure~\ref{fig:main graph} (top), we collect raw SQL queries $\tilde{q}$ from SQL performance discussions on community platforms such as StackOverflow. Because these queries are often written against incomplete or implicit database contexts, we adapt them to concrete schemas and dialects in our controlled environment. This process ensures reproducible execution and produces the executable source SQL $q_0$.
\paragraph{Fast SQL candidate generation.}
Given a source SQL $q_0$ and its database context $\mathcal{X}$, we generate 6 optimization candidates $\{q_1, \dots, q_6\}$. To ensure diverse coverage of the optimization space rather than collapsing to a single rewrite trajectory, we adopt a two-step process:
(1) \textbf{Heuristic summarization.} During crawling, we prompt an LLM to summarize recurring optimization patterns into 7 heuristic categories, requiring no manual curation.
(2) \textbf{$\epsilon$-greedy Candidate Generation.} 5 candidates are each guided by a composite focus drawn from the summarized heuristics, targeting distinct optimization directions. The remaining candidate is left unconstrained, allowing the LLM to freely explore optimization strategies beyond the predefined heuristics based on its own knowledge. The complete prompt templates are provided in Appendix~\ref{app:prompt}.
\paragraph{Pairwise SQL construction.}
We first execute each SQL in the candidate pool and discard queries that are non-executable or return results different from the source SQL $q_0$. Among the remaining semantically equivalent queries, we enumerate all pairwise combinations and order each pair by measured execution time, denoting the slower query as $q_i^{-}$ and the faster one as $q_i^{+}$. To avoid unstable pairs whose speed gap may disappear under execution noise, we retain a pair only when it achieves at least a $1.2\times$ measured speedup, a threshold justified in Appendix~\ref{app:speedup_threshold}, and exhibits at least a 20\% reduction in optimizer-estimated cost with non-trivial plan-level structural changes. These filters remove marginal or superficial edits and yield the final pair set:
\[
   \mathcal{P} = \{(q_i^{-}, q_i^{+})\}_{i=1}^{N},
\]
where $q_i^{-}$ and $q_i^{+}$ denote the slower and faster SQL in the $i$-th pair, respectively.

\subsection{Effi-SQL benchmark.}
\label{sec:Effi-SQL benchmark}
For evaluation, we further construct \textbf{Effi-SQL}, a benchmark annotated by human experts, to support reliable performance comparison.
\paragraph{Optimization annotation.}
Starting from the same source SQL collection method described in Section~\ref{sec:Automated scaffold}, human experts manually analyze the QEP of each source SQL, identify performance bottlenecks, and write an optimized SQL. Each optimized SQL is executed on the database engine to confirm semantic equivalence and genuine speedup.

\paragraph{Validation.}
After annotation, all pairs undergo cross-validation, with annotators exchanging data for independent review. Pairs involving ambiguous semantics or trivial optimizations are filtered out. The final benchmark achieves an inter-annotator agreement rate of 93.4\%. Full annotation guidelines and validation protocols are provided in Appendix~\ref{app:benchmark_construction}.

\paragraph{Benchmark statistics.}
Effi-SQL contains 1{,}100 Slow-Fast pairs across five SQL dialects: 300 PostgreSQL pairs and 200 pairs each for MySQL, SQL Server, Oracle, and SQLite. The benchmark is partitioned into three difficulty tiers — Simple, Moderate, and Challenging. The details of the benchmark are provided in Appendix~\ref{app:difficulty}.

\section{Diff-SQL: A two-stage patch-based framework for SQL optimization}
\label{sec:Diff-SQL}

\paragraph{Overview.}
Direct full query rewriting entangles efficiency optimization with correctness preservation in a single generation step. This induces \textbf{\textit{Objective Misalignment}} introduced in Section~\ref{sec:introduction}. To mitigate this tension, Diff-SQL formulates SQL optimization as a two-stage process in a unified diff patch space, consisting of the Patch Generator (Section~\ref{sec:patch_generator}) and the Constraint Aligner (Section~\ref{sec:constraint_aligner}), as shown in the lower panel of Figure~\ref{fig:main graph}.

\subsection{Diff-SQL scaffold with unified diff patches}
\label{sec:diff_sql_scaffold}

Diff-SQL first serves as a prompting scaffold for frontier LLMs. Instead of asking the model to regenerate the entire SQL query, we represent each optimization as a unified diff patch. This design follows software engineering practice, where patches expose only the edited spans while preserving the surrounding context of the original program~\citep{piterbarg2025d}. For SQL optimization, this patch space makes the intended edits explicit and anchors unchanged query regions to the original SQL.

Built on this representation, Diff-SQL decomposes optimization into two stages. The Patch Generator first analyzes the execution context and predicts where and how to edit the query, producing both an optimization reasoning trace and an initial patch. The Constraint Aligner then uses this reasoning trace as the intended optimization direction and revises the patch under executability and semantic-equivalence constraints. This inference-only instantiation is denoted as \textbf{Diff-ICL}.

\subsection{Patch Generator}
\label{sec:patch_generator}

\paragraph{Workflow.}
Given a slow SQL query $q$ and its input context $\mathcal{X}$, the Patch Generator outputs $(r,\hat{p})$, where $r$ is an optimization reasoning trace and $\hat{p}$ is the initial diff patch. The reasoning trace diagnoses the performance bottleneck and describes the intended edit plan, which both guides patch generation and provides the Constraint Aligner with the optimization intent to preserve.

\paragraph{Training for local models.}
To transfer this capability to smaller local models (SLMs), we perform supervised fine-tuning of the Patch Generator through optimization-distillation since SLMs require broader knowledge~\citep{DBLP:conf/acl/HsiehLYNFRKLP23,DBLP:conf/nips/WangW0B0020}. For each Slow-Fast pair $(q_i^-, q_i^+)$ constructed in Section~\ref{sec:pairs construction}, we derive a patch $p_i$ with a rule-based diff extractor and prompt a stronger teacher model to generate the corresponding reasoning trace $r_i^{*}$. This yields training triplets $(q_i^-, p_i, r_i^{*})$.

We further verify the faithfulness of the reasoning through a round-trip check: given only $(q_i^-, \mathcal{X}_i, r_i^{*})$, the teacher must regenerate a patch that preserves semantics, changes the execution plan, and achieves measurable speedup. From $5{,}213$ verified ordered pairs, this process yields $4{,}283$ validated PostgreSQL training instances, with a retention rate of 82.16\%.

\subsection{Constraint Aligner}
\label{sec:constraint_aligner}

\paragraph{Workflow.}
The initial patch from the Patch Generator may still be malformed, non-executable, or semantically inconsistent. The Constraint Aligner repairs these failures while preserving the intended optimization strategy. Given $(q,\mathcal{X},r,\hat{p})$, it outputs $(a,\tilde{p})$, where $a$ is a revision analysis and $\tilde{p}$ is the corrected patch. This stage focuses on enforcing patch validity, executability, and semantic equivalence rather than exploring a new optimization direction.

\paragraph{Training for local models.}
We train the Constraint Aligner with supervised warmup followed by execution-grounded reinforcement learning. In the warmup phase, we run the Patch Generator on the training set, execute the patched queries, and collect real execution errors or semantic mismatches. A teacher model then revises the failing patches, and only revisions that pass execution-based verification are retained as supervised data.

Warmup alone cannot cover the diverse failures encountered at inference time, so we further optimize the aligner with database feedback. For each input $(q,\mathcal{X},r,\hat{p})$, the model samples a group of $G$ revised outputs $\{(a_i,\tilde{p}_i)\}_{i=1}^{G}$. Each revised patch is applied to the original query and executed against the database to obtain a scalar reward $R_i$. We adopt GRPO~\citep{DBLP:journals/corr/abs-2402-03300,guo2025deepseek}, which computes advantages within each sampled group without requiring a separate critic:

\begin{equation}
\mathcal{J}_{GRPO}(\theta) = \mathbb{E} \left[
\frac{1}{G} \sum_{i=1}^{G} \left(
\min\left(
\rho_i A_i,\;
\text{clip}\left(\rho_i,\, 1{-}\varepsilon,\, 1{+}\varepsilon\right) A_i
\right)
- \beta\, \mathbb{D}_{KL}(\pi_\theta \| \pi_{ref})
\right)
\right],
\end{equation}

where each output $o_i=(a_i,\tilde{p}_i)$ contains a revision analysis and a revised patch, $\rho_i=\frac{\pi_\theta(o_i|q)}{\pi_{\theta_{old}}(o_i|q)}$ is the importance ratio, $A_i$ is the group-normalized advantage, and $\pi_{ref}$ is the reference policy. We use group size $G=8$ during training.

We also design a reward function which can distinguish fine-grained execution outcomes, including semantic match, timeout, system error, no-op or revert behavior, semantic mismatch, malformed patch, and execution error. This design encourages executable and semantically consistent repairs while discouraging trivial reverts. Full details of the reward design and efficient execution setup are provided in Appendix~\ref{app:reward}. The trained two-stage instantiation is denoted as \textbf{Diff-RL}.

\section{Experiment}
\label{sec:experiment}
\subsection{Experiment setup}

\paragraph{Evaluation metrics.}
We evaluate with two metrics. 
(1) \textbf{$\Delta$EX (Execution Accuracy Degradation)} measures the correctness loss introduced by optimization. Since all input queries are executable, we define $\Delta\mathrm{EX}=1-\mathrm{EX}$, where EX is the percentage of optimized queries that remain executable and semantically equivalent to the original. Lower $\Delta$EX indicates fewer optimization-induced correctness failures. (2) \textbf{R-VES (Reward-based Valid Efficiency Score)} additionally rewards execution speedup by weighting each matched SQL with a discretized gain function $g(\tau)$ based on the speedup ratio $\tau$, thereby reducing the impact of extreme cases to global overall experiments. To prevent database runtime noise from inflating efficiency scores, we adjust the $g(\tau)$ thresholds in R-VES so that meaningful speedup reward ($g(\tau) \ge 0.75$) requires $\tau \ge 1.2$. The full definition of the two metrics is provided in Appendix~\ref{app:preliminaries_formal}.

\paragraph{Evaluation database and scale.} We evaluate on four databases: \texttt{TPC-H} as a synthetic business database, together with three databases from BIRD-Interact~\citep{huo2026birdinteract} that cover different domains: \texttt{polar\_equipment}, \texttt{solar\_panel}, and \texttt{robot\_fault\_prediction}. For \texttt{TPC-H}, we use the 3GB scale factor. Although 10GB better approximates production workloads, its higher execution time makes repeated evaluation across multiple models prohibitively expensive. We validate in Appendix~\ref{app:scale_stability} that 3GB closely approximates 10GB in speedup ranking while keeping evaluation cost manageable. To stress-test optimization under heavier workloads, we scale the three BIRD-Interact databases to 10$\times$ their original size by proportionally replicating record tables while keeping feature tables unchanged. \paragraph{Baselines and settings.} We compare four settings. (1) \textbf{Human Designed Rewrite:} \texttt{R-Bot}~\citep{DBLP:journals/pvldb/SunZLYFZ25} uses LLM to select and order predefined rewrite rules, applies these rules through a query rewrite engine to produce the optimized SQL. (2) \textbf{Direct Full SQL Generation:} frontier LLMs (\texttt{GLM-4.7}~\citep{zai2026glm47}, \texttt{DeepSeek-V3}~\citep{DBLP:journals/corr/abs-2412-19437}, \texttt{MiniMax-M2}~\citep{minimax2025m2}, \texttt{Kimi-K2-Thinking}~\citep{team2025kimi}) directly generate optimized SQL from the input SQL. (3) \textbf{Diff-ICL:} We apply our two-stage pipeline Diff-ICL at inference time without any weight updates, evaluated on \texttt{Claude-Opus-4.5}~\citep{anthropic2025claudeopus45}, \texttt{Claude-Opus-4.6}~\citep{anthropic2026claudeopus46}, and \texttt{Qwen3-Coder}~\citep{yang2025qwen3}. (4) \textbf{Diff-RL:} We train the full pipeline on a 7B-scale model \texttt{Qwen-2.5-7B-Instruct}~\citep{qwen25}, motivated by deployment scenarios where data privacy constraints require local inference rather than relying on external APIs. 
\begin{table}
\caption{
Main results on the PostgreSQL subset of Effi-SQL. We compare \textbf{human-designed rewrite}, \textbf{direct full query generation}, \textbf{Diff-ICL}, and \textbf{Diff-RL}. 
\textbf{Lower $\Delta$EX} is better, and \textbf{higher R-VES} is better. Parenthesized values denote changes over the corresponding base model.
}
\centering
\scriptsize
\setlength{\tabcolsep}{2pt}
\hspace*{-0.02\linewidth}
\resizebox{1.0\linewidth}{!}{%
\begin{tabular}{l L{1.13cm} L{1.3cm} L{1.26cm} L{1.3cm} L{1.26cm} L{1.3cm} L{1.26cm} L{1.3cm}}
\toprule
& \multicolumn{2}{c}{\textbf{Simple}} 
& \multicolumn{2}{c}{\textbf{Moderate}} 
& \multicolumn{2}{c}{\textbf{Challenging}} 
& \multicolumn{2}{c}{\textbf{Overall}} \\
\cmidrule(lr){2-3}\cmidrule(lr){4-5}\cmidrule(lr){6-7}\cmidrule(lr){8-9}
\textbf{Model} 
& \textbf{$\Delta$EX(\%)}$\downarrow$ & \textbf{R-VES(\%)}$\uparrow$ 
& \textbf{$\Delta$EX(\%)}$\downarrow$ & \textbf{R-VES(\%)}$\uparrow$
& \textbf{$\Delta$EX(\%)}$\downarrow$ & \textbf{R-VES(\%)}$\uparrow$ 
& \textbf{$\Delta$EX(\%)}$\downarrow$ & \textbf{R-VES(\%)}$\uparrow$ \\
\midrule
\rowcolor{groupgray}
\multicolumn{9}{c}{\textit{\textbf{Human Designed Rewrite}}} \\
R-Bot        & 45.56 & 25.28 & 45.93 & 24.44 & 61.33 & 16.67 & 49.67 & 22.75 \\
\specialrule{0.35pt}{3pt}{2pt}
\rowcolor{groupgray}
\multicolumn{9}{c}{\textit{\textbf{Direct Full SQL Generation}}} \\
GLM-4.7                & 10.00 & 78.89 & 17.04 & 64.63 & 41.33 & 39.67 & 21.00 & 62.67 \\
DeepSeek-V3            & 5.56 & 77.22 & 17.04 & 61.48 & 32.00 & 45.67 & 17.33 & 62.25 \\
MiniMax-M2             & 24.44 & 66.39 & 37.04 & 48.33 & 50.67 & 31.33 & 36.67 & 49.50 \\
Kimi-K2-Thinking       & 35.56 & 59.72 & 44.44 & 45.37 & 49.33 & 35.67 & 43.00 & 47.25 \\

\specialrule{0.35pt}{3pt}{2pt}
\rowcolor{groupgray}
\multicolumn{9}{c}{\textit{\textbf{In-Context Learning}}} \\
Claude-Opus-4.5        & \underline{1.11} & 76.11 & \underline{1.48} & 74.63 & 20.00 & 59.67 & 6.00 & 71.33 \\
\hspace{0.8em} + Diff-ICL
& \textbf{0.00}\loss{-1.11}
& \underline{90.28}\gain{+14.17}
& \textbf{0.00}\loss{-1.48}
& \textbf{81.85}\gain{+7.22}
& \textbf{12.00}\loss{-8.00}
& \textbf{66.33}\gain{+6.66}
& \textbf{3.00}\loss{-3.00}
& \textbf{80.50}\gain{+9.17} \\

Claude-Opus-4.6        & \underline{1.11} & 70.83 & 5.19 & 69.63
 & 26.67 & 51.67 & 9.33 & 65.50 \\
\hspace{0.8em} + Diff-ICL
& \underline{1.11}\loss{-0.00}
& \textbf{90.56}\gain{+19.73}
& 2.22\loss{-2.97}
& \underline{77.96}\gain{+8.33}
& \underline{14.67}\loss{-12.00}
& \underline{62.67}\gain{+11.00}
& \underline{5.00}\loss{-4.33}
& \underline{77.92}\gain{+12.42}
\\

Qwen3-Coder            & 5.56 & 77.78 & 22.22 & 59.07 & 56.00 & 21.00 & 25.67 & 55.17 \\
\hspace{0.8em} + Diff-ICL
& \textbf{0.00}\loss{-5.56}
& 81.39\gain{+3.61}
& 7.41\loss{-14.81}
& 68.52\gain{+9.45}
& 45.33\loss{-10.67}
& 33.33\gain{+12.33}
& 14.67\loss{-11.00}
& 63.58\gain{+8.41}
\\

\specialrule{0.35pt}{3pt}{2pt}
\rowcolor{groupgray}
\multicolumn{9}{c}{\textit{\textbf{Two-Stage Training}}} \\
Qwen-2.5-7B-Instruct   & 34.44 & 49.72 & 43.70 & 30.00 & 54.67 & 20.00 & 43.67 & 33.42 \\

\hspace{0.8em} + Diff-RL
& 5.56\loss{-28.88}
& 75.00\gain{+25.28}
& 29.63\loss{-14.07}
& 42.04\gain{+12.04}
& 54.67\loss{-0.00}
& 21.67\gain{+1.67}
& 28.67\loss{-15.00}
& 46.83\gain{+13.41}
\\
\bottomrule
\end{tabular}%
}
\label{tab:main_results}
\end{table}
\subsection{Main results}
Table~\ref{tab:main_results} reports the main results in terms of $\Delta$EX and R-VES on the PostgreSQL subset of Effi-SQL. We next highlight four main observations. Details of the difficulty-level division are provided in Appendix~\ref{app:difficulty}.

\textbf{SQL optimization tends to introduce execution accuracy degradation.} 
While all methods improve efficiency to varying degrees, the optimization process introduces significant execution degradation. For example, \texttt{Kimi-K2-Thinking} incurs a $\Delta$EX of 43.00, meaning nearly half of the rewritten queries fail to remain correct. Because executability and semantic correctness are fundamental
requirements for SQL tasks, such a high degradation rate is unacceptable in practice. This partly reflects the \textbf{\textit{Objective Misalignment}}. Diff-SQL mitigates this by decoupling optimization from constraint enforcement. Both Diff-ICL and Diff-RL consistently improve R-VES while simultaneously reducing $\Delta$EX, demonstrating that the decomposed design can effectively alleviate this tension.

\textbf{Human-designed rewrite rules have limited generalization.} Most LLM-based methods outperform the rule-guided baseline R-Bot across both $\Delta$EX and R-VES. This is likely because human-predefined rewrite rules are inherently shaped by expert observation and prior bias, making them difficult to generalize to unseen or structurally diverse queries~\citep{DBLP:journals/pvldb/SunZLYFZ25,DBLP:journals/pvldb/LiYWCB24}. This limitation is amplified on Effi-SQL, where the optimization problems span diverse SQL structures and require a wide range of strategies beyond any fixed rule set. In contrast, LLM-based approaches analyze each query's execution context and can flexibly compose optimization strategies, enabling broader coverage over complex patterns.

\textbf{Diff-SQL is general and flexible.} (1) \textbf{As an inference-only framework,}
Diff-ICL applies the same pipeline at inference time without weight updates. Across all three base models, it consistently reduces $\Delta$EX and improves R-VES, with the largest gain on \texttt{Qwen3-Coder} ($-$11.00 $\Delta$EX, +8.41 R-VES). This shows that the decomposition is not tied to a specific model or training procedure, making Diff-SQL practical for production settings where additional training is costly or unavailable.
(2) \textbf{As a trainable framework for local deployment,}
Diff-RL reduces $\Delta$EX from 43.67 to 28.67 and improves R-VES from 33.42 to 46.83 on \texttt{Qwen-2.5-7B-Instruct}. These gains show that the same decomposition can also be learned by a 7B-scale local model, which is important for privacy-sensitive settings where schemas, query logs, and database contents must remain within the local environment.

\begin{wrapfigure}{r}{0.4\linewidth}
\centering
\includegraphics[draft=false,width=\linewidth]{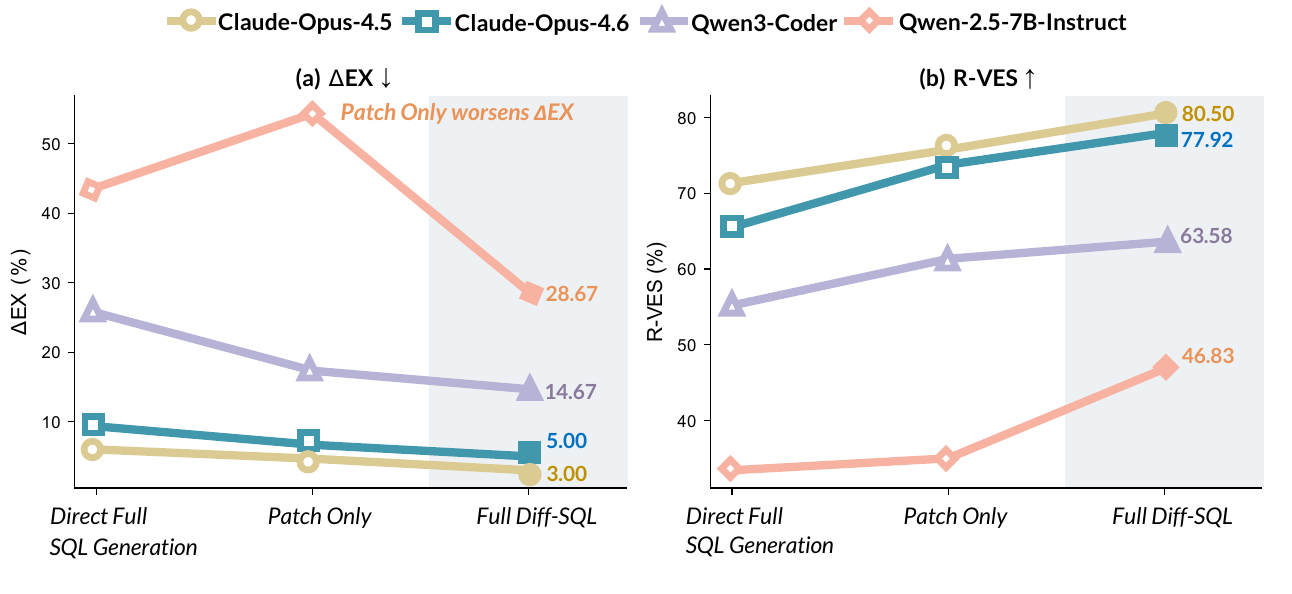}
\vspace{-4mm}
\caption{Ablation study of Diff-SQL.}
\label{fig:diffsql_design_analysis}
\end{wrapfigure}

\subsection{Ablation study}
To better understand where the gains come from, we next analyze the main design choices in Diff-SQL, as shown in Figure~\ref{fig:diffsql_design_analysis}.

\textbf{Patch-based editing benefits strong models but requires alignment for smaller models.}
Compared with direct full SQL rewriting, the Patch Generator reduces $\Delta$EX and improves R-VES on all three frontier models. However, when training only the Patch Generator on the 7B-scale model, $\Delta$EX increases from 43.67\% to 54.33\%, indicating that smaller models themselves struggle to produce structurally valid patches without downstream correction. 

\textbf{Refinement is necessary for reliable constraint alignment.}
We observe that adding the Constraint Aligner further improves performance on both metrics over patch-only (Patch Generator) prompting across all three models. The effect is more pronounced for smaller model, where Diff-RL reduces $\Delta$EX by 25.55\% and improves R-VES by 11.83\%. This indicates that the verification step mainly helps correct patch-level executability and semantic errors that are difficult to avoid through patch generation alone, making the optimized SQL more reliable, which in turn supports stronger efficiency gains. 

\textbf{Execution-grounded RL is essential for the Constraint Aligner.}
Under a fixed Patch Generator, SFT warmup alone yields only marginal improvement on the 7B model ($\Delta$EX: 54.33 $\to$ 51.67, R-VES: 35.00 $\to$ 36.00), showing that supervised imitation provides limited coverage of real execution failures. Adding execution-grounded RL dramatically improves both metrics ($\Delta$EX: 51.67 $\to$ 28.67, R-VES: 36.00 $\to$ 46.83). This confirms that the Patch Generator can already propose useful optimization directions, but the Constraint Aligner requires RL with deterministic database feedback to consistently turn these proposals into correct and efficient SQL.

\section{What Shapes SQL Optimization Performance}
\vspace{-1 mm}
To better understand what drives optimization quality, we analyze outputs from three complementary views. Section~\ref{sec:reasoning} studies reasoning traces, Section~\ref{sec:patch} studies patch size and rewrite scope, and Section~\ref{sec:error} studies failure categories across all evaluated settings. 
\vspace{-1 mm}
\subsection{Reasoning analysis}
\label{sec:reasoning}

\paragraph{Optimization operators and semantic guards are complementary.}
Our prompt requires both speedup and semantic equivalence, so models naturally produce two kinds of reasoning: \textbf{Optimization operators} that identify a concrete transformation and \textbf{Semantic guards} that verify equivalence invariants.
Operator-only reasoning proposes transformations but is semantically fragile, resulting in the largest EX degradation among the reasoning categories at 36\% and an R-VES of 44.2\%. 
Semantic-only reasoning has the opposite profile: it is safer, with the lowest $\Delta$EX of 12.2\%, but more conservative, with an R-VES of 50.3\%. 
The best overall trade-off appears when both components are present, achieving the highest R-VES of 56.4\% while maintaining a strong $\Delta$EX of 22.1\%. 
This suggests that the combination of optimization operators and semantic guards is important for reliable SQL optimization.

\paragraph{Longer reasoning is not always better.}
Reasoning traces under 120 tokens achieve an R-VES of 82.1\%, whereas traces longer than 451 tokens drop to 51.4\%. We hypothesize that this decline reflects \textit{hesitation} because long traces often contain repeated self-revision (60.0\%), where the model proposes, revises, and replaces its own SQL candidates within a single response. Such behavior suggests that the model has not formed a stable optimization decision; instead, later revisions may drift away from an earlier valid rewrite or introduce unnecessary structural changes. This pattern is most evident in \textit{Kimi-K2-Thinking}, which produces 8.18 SQL blocks on average. Although its successful rewrites achieve a high matched speedup of 2.534, the frequent candidate switching makes its final optimization behavior less reliable.

\subsection{Patch and rewrite-scope analysis}
\label{sec:patch}
\paragraph{Moderate patch size performs best.}
We further measure patch size by the number of physical changed lines. For Diff-ICL, patches with 3--5 changed lines achieve the highest R-VES, while performance declines once patches exceed 11 changed lines. The safe range is narrower for the 7B model. When patches contain 21 or more changed lines, R-VES drops sharply to 14.8\%. This indicates that effective SQL optimization usually requires localized but non-trivial edits, whereas overly large patches increase the risk of semantic drift and execution failure.
\begin{wrapfigure}[9]{l}{0.45\linewidth}
\vspace{10pt}
\centering
\includegraphics[draft=false,width=\linewidth]{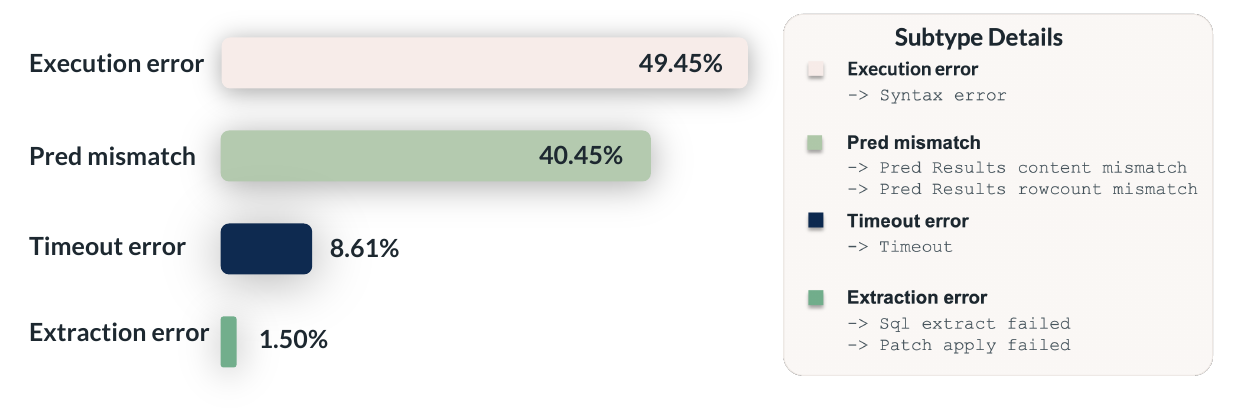}
\vspace{-10pt}
\caption{Error category distribution.}
\label{fig:error_category_bar_mapping_native}
\vspace{-5 mm}
\end{wrapfigure}
\paragraph{Changed clause count.}
While line-level patch size captures textual edit scope, it does not reflect structural edit scope. We compute the changed clause count from the source SQL and the final optimized SQL, regardless of whether the method internally generates a full query or a patch. This metric counts modifications across seven high-level SQL regions: \texttt{CTE}, \texttt{SELECT}, \texttt{FROM/JOIN}, \texttt{WHERE}, \texttt{GROUP BY/HAVING}, \texttt{ORDER BY/LIMIT}, and \texttt{WINDOW}. Under this measure, patch-based optimization yields more bounded rewrites than direct full query generation. Direct generation exhibits a broad-rewrite tail, with 10.77\% of outputs modifying 6 or more regions, where $\Delta$EX is higher and R-VES is lower than in the $\leq$5-region bucket. In contrast, Diff-ICL keeps 99.53\% of rewrites within at most 5 changed regions while achieving stronger overall scores in this bounded range. This suggests that patching does not merely reduce textual edit size; it anchors optimization to a limited set of relevant SQL clauses and avoids unnecessary broad regeneration.

\subsection{Error distribution}
\label{sec:error}

As shown in Figure~\ref{fig:error_category_bar_mapping_native}, the dominant error across all settings is \texttt{execution\_error}, where the optimized SQL or Patch fails to execute. It accounts for 49.45\% of all failures. The second most frequent error is \texttt{pred\_mismatch}, where the rewrite executes successfully but returns different values—typically caused by missing semantic guards on NULLs, ties, duplicates, or aggregation grain. It accounts for 40.45\% of all failures. Together, these two categories cover the majority of failures and reflect the same underlying issue: objective misalignment during SQL optimization. These patterns indicate that the main bottleneck is not merely proposing an optimization direction, but realizing it as an executable and semantically equivalent SQL rewrite.

\vspace{-1 mm}
\section{Related Work}
\vspace{-1 mm}
\textbf{Text-to-SQL.}\quad Text-to-SQL has become a central problem in database reasoning, with benchmarks such as Spider~\citep{DBLP:conf/emnlp/YuZYYWLMLYRZR18,DBLP:conf/iclr/LeiCYCSSSGHYZX025} and BIRD~\citep{li2023can,livesqlbench2025,li2026swesql,huo2026birdinteract} driving progress on semantic correctness, schema grounding, and execution accuracy. Recent methods span in-context decomposition and prompting, data-centric training, and agent-based interaction, exemplified by DIN-SQL~\citep{DBLP:conf/nips/PourrezaR23} and DAIL-SQL~\citep{DBLP:journals/pvldb/GaoWLSQDZ24}, CodeS~\citep{DBLP:journals/pacmmod/LiZLFZZWP0024} and DTS-SQL~\citep{DBLP:conf/emnlp/PourrezaR24}, and MAC-SQL~\citep{DBLP:conf/coling/WangR0LBCYZYSL25}, respectively. BIRD further emphasizes that generated SQL should be not only correct but also efficient, with the primary setting still centered on NL-to-SQL generation from user intent. This differs from our setting, which starts from an existing slow SQL query and focuses on optimizing it for efficiency.

\textbf{SQL rewriting and efficiency-aware code optimization.}\quad
Traditional SQL optimization relies on database query optimizers and rule-based rewrite systems that select transformations from predefined heuristic rules. Recent work has begun to investigate whether large language models can go beyond fixed optimizer rules to perform more flexible SQL rewrite optimization~\citep{DBLP:journals/pvldb/SunZLYFZ25,DBLP:journals/pvldb/LiYWCB24}. However, they remain constrained by the predefined rule space of the underlying optimizer. To explore the potential of more direct LLM-based rewriting, other approaches, including QUITE~\citep{DBLP:journals/corr/abs-2506-07675}, LITHE~\citep{DBLP:conf/edbt/DharwadaDHD26}, and E$^3$-Rewrite~\citep{DBLP:journals/corr/abs-2508-09023}, investigate database-aware prompting, agentic search, execution feedback, semantic verification, and learned rewriting objectives. Beyond SQL, code efficiency has also attracted growing attention. On the benchmarking side, EffiBench~\citep{DBLP:conf/nips/0005QSCZ24,DBLP:journals/corr/abs-2505-13004}, COFFE~\citep{DBLP:journals/pacmse/PengWLR25}, ECCO~\citep{DBLP:conf/emnlp/WaghjaleVWF24}, and SWE-Perf~\citep{DBLP:journals/corr/abs-2507-12415} evaluate time or resource efficiency across code generation, code editing, and repository-level settings, while on the method side, EffiLearner~\citep{DBLP:conf/nips/0005DWWQCG024} explicitly optimizes generated code guided by execution-time and memory profiles.

\textbf{Reinforcement learning for coding.}\quad Reinforcement learning has been used to improve code generation and software engineering with signals from testing, execution, compilation, and downstream software outcomes, as in CodeRL~\citep{le2022coderl}, PPOCoder~\citep{DBLP:journals/tmlr/ShojaeeJTR23}, StepCoder~\citep{DBLP:conf/acl/Dou0JZXSHWFXZJZ24}, and SWE-RL~\citep{DBLP:journals/corr/abs-2502-18449}. More recent work also begins to explore training via GRPO, popularized in broader reasoning settings by DeepSeek-R1~\citep{guo2025deepseek}. RL has also been applied to tasks related to SQL, including Reasoning-SQL~\citep{DBLP:journals/corr/abs-2503-23157} and Arctic-Text2SQL-R1~\citep{DBLP:journals/corr/abs-2505-20315}. Our setting is more tightly execution-grounded, where GRPO serves to constrain and refine upstream optimization proposals into executable and semantically correct query revisions.
\vspace{-1 mm}
\section{Conclusion}
\vspace{-1 mm}
In this paper, we identify \textbf{\textit{Objective Misalignment}} as a key challenge in SQL efficiency optimization. To address this problem, we propose \textbf{Diff-SQL}, a two-stage framework that separates efficiency-oriented patch generation from constraint-aware alignment. Diff-SQL supports two instantiations as \textbf{Diff-ICL}, an inference-only prompting framework for frontier LLMs, and \textbf{Diff-RL}, an execution-grounded training framework for local models.
To obtain reliable optimization supervision, we introduce an automated pipeline to construct Slow-Fast SQL pairs. We further build \textbf{Effi-SQL}, a multi-dialect human-verified benchmark for standardized evaluation. Across both inference-only and trained settings, Diff-SQL improves the efficiency-correctness trade-off, showing that reliable SQL optimization requires explicit alignment between optimization intent, semantic preservation, and real execution behavior.
\newpage

\bibliographystyle{plainnat}
\bibliography{references}
\newpage
\appendix
\section{Experiment Details}
\label{app:experiment_details}

\subsection{API and Model Access}

For proprietary or closed-source models, we access model outputs through their official or supported API providers.
For open-source models, we use locally hosted checkpoints and perform inference with \texttt{vLLM}.
All prompts use the same input fields described in Appendix~\ref{app:prompt}, including the original SQL, schema information, column statistics, and query execution plan context when available.
Table~\ref{tab:api_model_access} summarizes the model access details.

\begin{table}[H]
\centering
\caption{Model access and API providers used in our experiments.}
\label{tab:api_model_access}
\small
\setlength{\tabcolsep}{6pt}
\renewcommand{\arraystretch}{1.08}
\begin{tabular}{lll}
\toprule
\textbf{Model family} & \textbf{Access} & \textbf{Provider / source} \\
\midrule
Claude-Opus models & Official API & Anthropic \\
DeepSeek models & Official API & DeepSeek Platform \\
GLM models & Official API & Zhipu AI / Z.AI \\
MiniMax models & Official API & MiniMax Platform \\
Kimi models & Official API & Moonshot AI / Kimi \\
Qwen3-Coder & Official API & Alibaba Cloud DashScope / Qwen \\
Qwen2.5-7B-Instruct & Local checkpoint & Qwen / Hugging Face \\
\bottomrule
\end{tabular}
\end{table}

\subsection{Data, Model, and License Information}

Table~\ref{tab:license_information} summarizes the external resources used in our experiments and their corresponding license or usage terms.
Raw crawling outputs, API-generated intermediate results, and trained model checkpoints are not included in the submitted package.

\begin{table}[H]
\centering
\caption{Data, model, and license information.}
\label{tab:license_information}
\small
\setlength{\tabcolsep}{5pt}
\renewcommand{\arraystretch}{1.12}
\begin{tabular}{p{0.24\linewidth}p{0.32\linewidth}p{0.36\linewidth}}
\toprule
\textbf{Resource} & \textbf{Usage in this work} & \textbf{License / usage note} \\
\midrule
StackOverflow / Stack Exchange posts
& Source for SQL performance discussions and optimization rationales
& Creative Commons Attribution-ShareAlike 4.0 (CC BY-SA 4.0) \\

BIRD-Interact databases
& Real-world-style database environments for data construction and evaluation
& Used following the released dataset terms; released under CC BY-SA 4.0 when distributed through the public dataset release \\

TPC-H
& Synthetic analytical workload and scalable database environment
& Used following the TPC benchmark specification and fair-use policy \\

Qwen2.5-7B-Instruct
& Backbone model for Diff-RL training
& Apache 2.0 license \\

API-generated model outputs
& Intermediate outputs for prompt-based construction and evaluation
& Generated through official model APIs; raw API outputs are not included in the submitted package \\
\bottomrule
\end{tabular}
\end{table}

\subsection{Training Infrastructure}

We train local models on 4 NVIDIA A100 GPUs.
Training is implemented with the \texttt{verl} framework for reinforcement learning and distributed execution.
Both SFT and RL training use LoRA adapters.
The training data are split into train and validation sets with a 9:1 ratio.
Table~\ref{tab:training_infrastructure} summarizes the general training setup.

\begin{table}[H]
\centering
\caption{Training infrastructure and general setup.}
\label{tab:training_infrastructure}
\small
\setlength{\tabcolsep}{7pt}
\renewcommand{\arraystretch}{1.1}
\begin{tabular}{ll}
\toprule
\textbf{Item} & \textbf{Setting} \\
\midrule
GPU hardware & 4$\times$ NVIDIA A100 80G \\
Training framework & \texttt{verl} \\
Backbone model & Qwen2.5-7B-Instruct \\
Adapter method & LoRA \\
Inference engine & \texttt{vLLM} \\
RL reward function & \texttt{sql\_optimize} \\
\bottomrule
\end{tabular}
\end{table}

\subsection{SFT Hyperparameters}

We perform supervised fine-tuning for both the Patch Generator and the Constraint Aligner.
Table~\ref{tab:sft_hyperparameters} lists the main SFT hyperparameters.

\begin{table}[H]
\centering
\caption{SFT hyperparameters.}
\label{tab:sft_hyperparameters}
\small
\setlength{\tabcolsep}{7pt}
\renewcommand{\arraystretch}{1.08}
\begin{tabular}{ll}
\toprule
\textbf{Hyperparameter} & \textbf{Value} \\
\midrule
\texttt{lora\_rank} & 64 \\
\texttt{lora\_alpha} & 128 \\
\texttt{total\_epochs} & 4 \\
\texttt{learning rate} & $5\times10^{-5}$ \\
\texttt{weight\_decay} & 0.01 \\
\bottomrule
\end{tabular}
\end{table}

\subsection{RL Hyperparameters}

For the Constraint Aligner, we further apply GRPO style reinforcement learning with an execution grounded reward.
The reward function validates patch extraction, patch application, SQL executability, and semantic equivalence by executing the patched query against a real database.
Table~\ref{tab:rl_hyperparameters} lists the main RL hyperparameters.

\begin{table}[H]
\centering
\caption{GRPO hyperparameters for the Constraint Aligner.}
\label{tab:rl_hyperparameters}
\small
\setlength{\tabcolsep}{7pt}
\renewcommand{\arraystretch}{1.08}
\begin{tabular}{ll}
\toprule
\textbf{Hyperparameter} & \textbf{Value} \\
\midrule
\texttt{total\_steps} & 300 \\
\texttt{actor.lr} & $5\times10^{-6}$ \\
\texttt{rollout.n} & 8 \\
\bottomrule
\end{tabular}
\end{table}

\subsection{RL Execution Environment}

RL training requires executing each sampled patch against a real database, making execution throughput a practical bottleneck.
Since the Constraint Aligner reward is based on patch validity, executability, and semantic equivalence rather than absolute execution latency, we conduct RL exploration on smaller databases to reduce feedback latency.
During RL exploration, we run TPC-H instances on the 0.1G scale database and BIRD-Interact instances on their original unscaled databases.
This setting allows the RL loop to obtain faster database feedback while still checking whether the patched query executes and preserves the original result set.
Small-scale execution is used only for RL feedback collection, not for reporting final efficiency results.
Final evaluation, including EX, R-VES, and latency-based speedup, is performed on the full evaluation scale.

\begin{table}[H]
\centering
\caption{Database scale used for RL feedback collection and final evaluation.}
\label{tab:rl_eval_database_scale}
\small
\setlength{\tabcolsep}{7pt}
\renewcommand{\arraystretch}{1.1}
\begin{tabular}{lll}
\toprule
\textbf{Workload} & \textbf{RL feedback collection} & \textbf{Final evaluation} \\
\midrule
TPC-H & 0.1G scale database & Full evaluation scale \\
BIRD-Interact & Original unscaled databases & Full evaluation scale \\
\bottomrule
\end{tabular}
\end{table}

\subsection{Training Cost}

Table~\ref{tab:training_cost} reports the approximate training time on our 4$\times$A100 setup.
The estimates include training time only and exclude data collection, database construction, API inference, and final execution evaluation.

\begin{table}[H]
\centering
\caption{Approximate training cost on 4$\times$A100 GPUs.}
\label{tab:training_cost}
\small
\setlength{\tabcolsep}{8pt}
\renewcommand{\arraystretch}{1.1}
\begin{tabular}{lll}
\toprule
\textbf{Stage} & \textbf{Training objective} & \textbf{Approx. time} \\
\midrule
Patch Generator + Constraint Aligner & SFT & 4.5 hours \\
Constraint Aligner & GRPO / RL & 30 hours \\
\bottomrule
\end{tabular}
\end{table}

\subsection{Execution Evaluation Protocol}
\label{app:execution_eval_protocol}

All reported execution metrics are obtained by executing SQL queries against real database instances.
For each evaluation example, we first validate semantic correctness by comparing the full result sets of the rewritten SQL and the original SQL.
The comparison uses row-level fingerprints and preserves duplicate counts, so two queries are considered matched only when they return the same multiset of rows.
If the row counts or row fingerprints differ, the prediction is marked as a semantic mismatch.
Runtime measurement is performed only for predictions that execute successfully and match the original query result.

For runtime measurement, we execute each matched query pair multiple times and use the median runtime to reduce execution noise.
Unless otherwise specified, each query pair is evaluated with three runs.
In each run, the rewritten SQL and original SQL are executed in a randomly shuffled order to reduce systematic ordering effects.
Let $T_i^{\mathrm{orig}}$ and $T_i^{\mathrm{rew}}$ denote the median execution times of the original and rewritten SQL for example $i$, respectively.
The measured speedup is computed as
\begin{equation}
    \mathrm{Speedup}_i = \frac{T_i^{\mathrm{orig}}}{T_i^{\mathrm{rew}}}.
\end{equation}
We cap extremely large ratios at 1000 to avoid numerical instability from near-zero runtimes.

We use a 300-second timeout for query execution.
For PostgreSQL and MySQL timing runs, execution time is obtained through native execution-based plan interfaces: \texttt{EXPLAIN (ANALYZE, FORMAT JSON)} for PostgreSQL and \texttt{EXPLAIN ANALYZE} for MySQL.
For other dialects, we execute the query and measure wall-clock runtime after fetching the result set.
All executions are rolled back after each run to avoid persistent side effects.
The evaluator records execution failures, timeouts, semantic mismatches, row counts, median runtimes, and speedup for downstream EX and R-VES computation.

\section{Benchmark Construction and Human Verification}
\label{app:benchmark_construction}
\begin{figure}
    \centering
    \includegraphics[draft=false,width=0.5\linewidth]{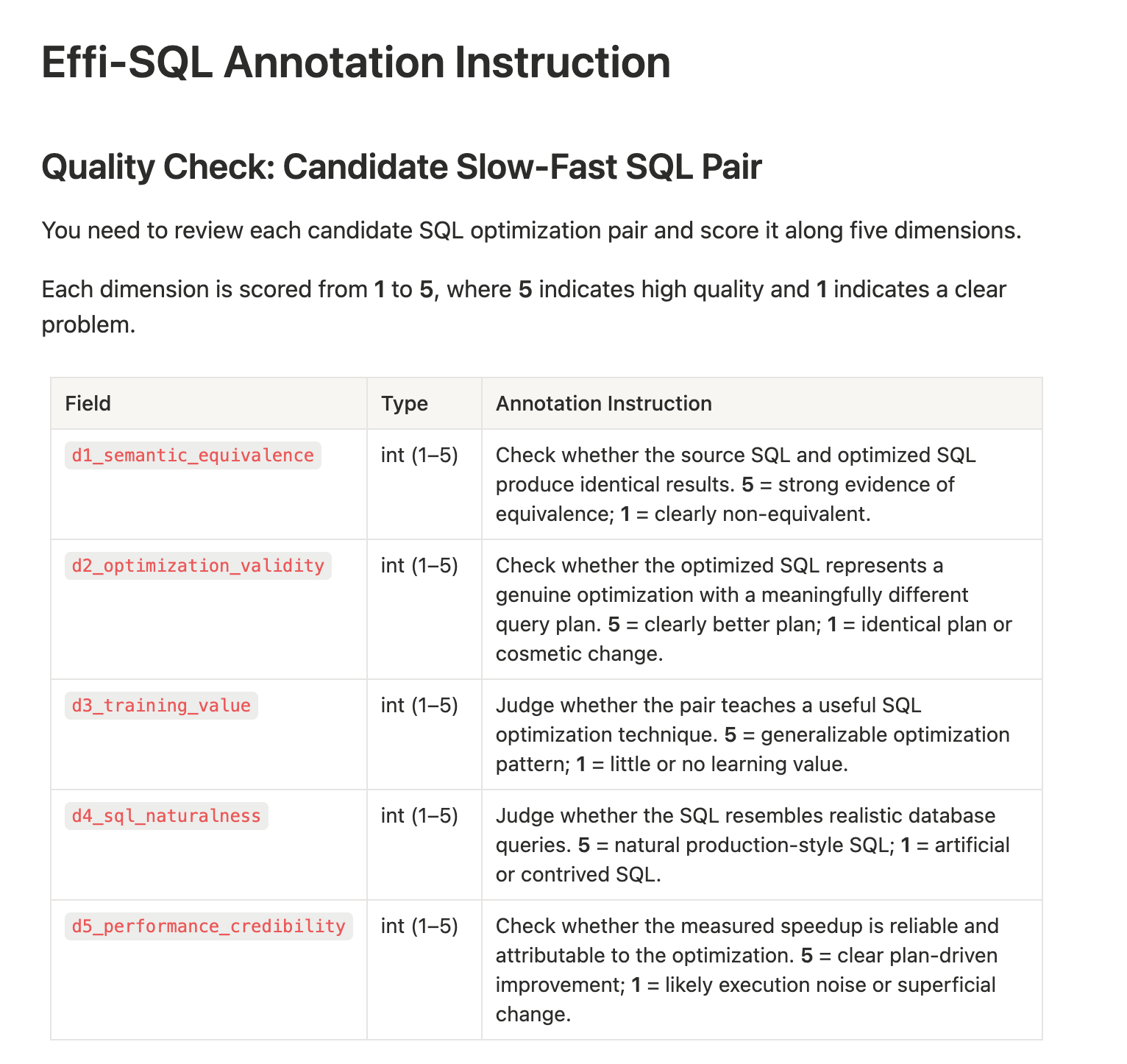}
    \caption{Example screenshot of the annotation instruction used for human verification.}
    \label{fig:annotation_screenshot}
\end{figure}
We construct a multi-dialect SQL optimization benchmark covering five SQL dialects.
We first crawl StackOverflow posts related to SQL performance optimization and collect the source SQL queries discussed in the posts.
The optimized SQL is not available a priori; instead, it is constructed by annotators through two routes.
When the original author or commenters provide a concrete optimization suggestion, annotators adapt the suggested rewrite to our target schema and dialect, execute it in our database environment, and verify its correctness and efficiency.
When the discussion does not provide a clearly effective rewrite, or when the suggested rewrite fails to yield reliable improvement, annotators inspect the corresponding query execution plan to identify the actual execution bottleneck and manually construct an optimized SQL query.
A candidate Slow-Fast pair is retained only when the optimized SQL is executable, result-equivalent to the source SQL, and shows a clear execution-time improvement.

\subsection{Annotation Platform and Compensation}

We build an internal annotation environment for SQL optimization construction and verification.
The platform provides annotators with database access, execution scripts, query plan extraction tools, and API-based assistance for inspecting SQL behavior, understanding StackOverflow discussions, and validating candidate rewrites.
Each participant is also provided with an API coupon to cover the cost of using model-assisted analysis tools during annotation.
Annotators are compensated according to instance difficulty: simple instances are compensated at 5 USD per validated instance, while difficult instances are compensated at 10 USD per validated instance.
The higher compensation for difficult instances reflects the additional effort required for interpreting optimization discussions, inspecting execution plans, checking semantic equivalence, and manually repairing or constructing optimized SQL queries.

\subsection{SQL Optimization Experts}

We invite SQL optimization experts to supervise the benchmark construction and verification process.
The expert group consists of researchers and practitioners with experience in database systems, SQL optimization, and execution-based evaluation.
Their responsibilities include reviewing ambiguous cases, checking whether an observed speedup reflects a genuine optimization, validating semantic-equivalence decisions, and resolving disagreements among annotators.
In particular, experts inspect cases involving complex joins, aggregation-grain changes, NULL-sensitive predicates, duplicate preservation, outer-join semantics, and dialect-specific SQL behavior.

\subsection{Annotator Qualification}

Annotators are required to have database-related background knowledge and practical experience with SQL.
Before participating in the main annotation process, each annotator completes a qualification test consisting of SQL optimization examples with known verification outcomes.
The test evaluates whether the annotator can: (1) execute and compare SQL queries in the provided database environment; (2) identify execution failures and semantic mismatches; (3) read query execution plans and locate major bottlenecks; and (4) judge whether a rewritten query preserves the original query semantics.
Only annotators who pass the qualification test are allowed to participate in benchmark construction and verification.

\subsection{Human Verification Procedure}

Each candidate Slow-Fast pair is reviewed along five quality dimensions: semantic equivalence, optimization validity, training value, SQL naturalness, and performance credibility.
For each instance, annotators inspect the source SQL, optimized SQL, schema context, execution results, and runtime measurements.
When necessary, they rerun the queries in the provided database environment, inspect query execution plans, and check whether the observed speedup reflects a genuine optimization rather than execution noise or superficial rewriting.

The verification process follows a multi-round protocol.
In the first round, annotators check executability and result equivalence through real database execution.
In the second round, they review optimization validity and remove cases where the optimized SQL is semantically fragile, dialect-invalid, or only superficially different from the source SQL.
In the third round, experts revisit ambiguous or disputed cases and either confirm removal, repair the SQL, or retain the pair after execution-based re-verification.
Only pairs that pass both automatic execution checks and human quality review are retained in the benchmark.

\subsection{Verification}

This process yields 3,065 candidate Slow-Fast SQL pairs.
Starting from these candidates, we conduct a multi-round verification process with human review.
Each pair is validated through real database execution and reviewed along the five quality dimensions described above.
After three rounds of quality review, 2,009 pairs are retained and 1,056 candidates are removed.
Among the removed instances, 986 cases (93.4\%) are explicitly confirmed for removal during second-pass review, while only 70 cases (6.6\%) are conservatively removed due to unresolved ambiguity.
We further revisit the removed set and recover 85 additional valid pairs through SQL repair and execution-based re-verification, resulting in 2,094 verified multi-dialect optimization pairs. 

\subsection{Final Benchmark Selection}

From this verified multi-dialect pool, we select 1,100 instances by jointly considering dialect coverage, difficulty distribution, structural redundancy, and query-pattern diversity. 
This final selection step is not intended to rank verified pairs by quality: all instances in the pool have already passed execution-based verification and human review. 
Instead, it is used to construct a balanced and non-redundant benchmark, avoiding over-representation of near-duplicate query structures or overly concentrated difficulty levels. 
The final benchmark contains 300 PostgreSQL instances and 200 instances each for MySQL, SQLite, MSSQL, and Oracle.

\section{Ethics}
\label{app:ethic}
This research complies with the NeurIPS Code of Ethics. We have carefully reviewed the guidelines and ensured that our work aligns with the stated ethical standards, including considerations of data privacy, fairness, and responsible use of released datasets and code.
This work involves crowdsourcing human annotation for SQL optimization benchmark construction. 
The annotation does not introduce bias. 
Annotation instructions, example screenshots, qualification details, and compensation information are provided in Appendix~\ref{app:benchmark_construction}.

\section{Difficulty Level}
\label{app:difficulty}
We assign benchmark instances into three difficulty levels: Simple, Moderate, and Challenging. The assignment is based on a hybrid difficulty score that combines empirical model performance with static SQL complexity. We use Qwen-2.5-7B-Instruct as a reference model and evaluate each instance by whether the model produces an executable and result-equivalent rewrite, together with the runtime-aware optimization score achieved by the rewrite. Instances that the reference model fails to solve, or solves with limited runtime improvement, are considered empirically harder.

Formally, we define the empirical difficulty as:
\[
S_{\mathrm{emp}} = 1 - \frac{\mathrm{EX} + \mathrm{R\text{-}VES}}{2}.
\]
Here, EX measures execution correctness, and R-VES measures runtime-aware optimization quality.

To avoid assigning difficulty solely based on one model's behavior, we also compute a static structural complexity score from the slow SQL query. This score aggregates normalized indicators such as query length, number of joins, number of subqueries, CTEs, window functions, set operations, aggregation, conditional expressions, DISTINCT, NULL-related predicates, and SQL function calls. We normalize this structural score across all benchmark instances.

The final hybrid difficulty score is:
\[
S_{\mathrm{hybrid}}
=
0.7 \cdot S_{\mathrm{emp}}
+
0.3 \cdot S_{\mathrm{struct}}.
\]
For the PostgreSQL subset used in the main experiments, we sort all instances by \(S_{\mathrm{hybrid}}\) and divide them into three groups, corresponding to Simple, Moderate, and Challenging. The resulting buckets contain 90, 135, and 75 instances in PostgreSQL subset, respectively.
\begin{table}[H]
\centering
\scriptsize
\caption{SQL length and structural complexity across difficulty levels of the PostgreSQL subset of Effi-SQL.}
\begin{tabular}{lrrrrrrrr}
\toprule
Difficulty & \#Inst. & Avg Len & Median Len & Max Len & Avg Tokens & Avg Joins & Avg Subq. & Avg Aggs \\
\midrule
Simple      & 90  & 212.70 & 194 & 542  & 30.60 & 0.40 & 0.97 & 0.80 \\
Moderate    & 135 & 305.87 & 287 & 875  & 44.04 & 0.76 & 1.41 & 0.89 \\
Challenging & 75  & 544.37 & 457 & 2424 & 75.31 & 1.63 & 2.12 & 2.03 \\
Overall & 300 & 337.55 & 271 & 2424 & 47.83 & 0.87 & 1.45 & 1.15 \\
\bottomrule
\end{tabular}
\label{tab:difficulty-stats}
\end{table}
\subsection{Database Scale Statistics}
\label{app:db_statistics}

In addition to SQL-level structural complexity, database scale also affects execution behavior.
Table~\ref{tab:db_statistics} reports the database-level statistics used in Effi-SQL evaluation.
These statistics are reported separately from the difficulty buckets, since difficulty assignment is based on model behavior and SQL structural complexity rather than database size.

\begin{table}[H]
\centering
\caption{Database scale statistics used in Effi-SQL evaluation.}
\label{tab:db_statistics}
\small
\setlength{\tabcolsep}{7pt}
\renewcommand{\arraystretch}{1.08}
\begin{tabular}{lrrr}
\toprule
\textbf{Database} & \textbf{\#Tables} & \textbf{Total Rows} & \textbf{Avg. Rows / Table} \\
\midrule
\texttt{polar\_equipment\_scale} & 16 & 1,085,946 & 67,871.62 \\
\texttt{solar\_panel\_scale} & 10 & 852,304 & 85,230.40 \\
\texttt{robot\_fault\_prediction\_scale} & 10 & 862,200 & 86,220.00 \\
\texttt{tpch\_3g} & 8 & 25,976,639 & 3,247,079.88 \\
\midrule
Overall & 44 & 28,777,089 & 654,024.75 \\
\bottomrule
\end{tabular}
\end{table}
\section{Benchmark comparison}
\begin{table}[H]
\caption{Data statistics of features in Effi-SQL compared to related benchmarks.}
\label{tab:benchmark_comparison}
\centering
\scriptsize
\setlength{\tabcolsep}{5pt}
\begin{tabular}{llllllll}
\toprule
\textbf{Benchmark} & \textbf{\# Instances} & \textbf{\# DBs} & \textbf{\# Dialects} & \textbf{\# Evaluation Metric} & \textbf{Avg Code Len} & \textbf{Input} & \textbf{Task} \\
\midrule
Spider         & 2,147 & 200 & 1 & EM/EX & 30.18  & NL          & Text-to-SQL \\
BIRD-SQL       & 1,789 & 15  & 1 & EX/VES & 50.01  & NL          & Text-to-SQL \\
BIRD-CRITIC    & 1,100 & 11  & 4 & Test Cases & 112.08 & Given SQL   & SQL debugging \\
EffiBench      & 1,000    & --  & 1 & Pass@1/Time/Mem. & 475.84     & Python        & Code efficiency \\
COFFE          & 756    & --  & 1 & efficient@k & 298.10     & Python        & Code efficiency \\
\textbf{Effi-SQL (Ours)} & 1,100 & 4 & 5 & EX/R-VES & 384.55 & Given SQL & SQL efficiency \\
\bottomrule
\end{tabular}
\end{table}
Table~\ref{tab:benchmark_comparison} compares Effi-SQL with representative benchmarks for Text-to-SQL, SQL debugging, and code efficiency. Existing Text-to-SQL benchmarks such as Spider and BIRD-SQL primarily evaluate whether a model can synthesize correct SQL from natural language questions. BIRD-CRITIC moves closer to execution-facing SQL tasks by evaluating SQL issue debugging from a given SQL input, but its goal is error repair rather than execution efficiency optimization. EffiBench and COFFE focus on code efficiency, but they are centered on Python programs rather than database queries. In contrast, Effi-SQL targets SQL efficiency optimization directly. Each instance starts from a given slow SQL query and evaluates whether the model can produce a semantically equivalent but faster rewrite under real database execution. Effi-SQL also covers five SQL dialects and uses EX/R-VES to jointly measure semantic correctness and runtime-aware optimization quality.

\section{Limitations and future work.}
\label{app:limitations}
This work focuses on SQL optimization, one representative instance of system-level execution optimization. Extending the framework to broader code-based optimization tasks such as C++ or CUDA performance tuning is our future direction.

In addition, our current evaluation relies on statically enlarged databases, which incur substantial execution latency and are not well suited for interactive, agentic optimization workflows. Although agentic interaction could provide more fine-grained optimization clues through iterative exploration and execution feedback, the high cost of repeated database queries makes this impractical under our current infrastructure. Designing lighter execution scaffolds or tool-augmented pipelines to enable efficient agentic interaction is a promising direction. Nonetheless, even without agentic exploration, our two-stage decomposition already alleviates objective misalignment and delivers strong improvements in both efficiency and correctness.

\section{Broader impact.}
\label{app:broder impact}
The objective misalignment identified in this work is a general phenomenon in multi-objective optimization. When a model must simultaneously maximize one objective while satisfying hard constraints on another, single-pass generation tends to over-commit to the primary objective at the expense of constraint satisfaction. The two-stage decomposition proposed here, separating objective pursuit from constraint alignment, may offer an applicable paradigm beyond SQL optimization. Furthermore, our training recipe for 7B-scale models is particularly relevant for database applications where data privacy requirements demand local deployment, providing a practical path toward execution-efficient SQL optimization without relying on external API services.

\section{Over rewrite on direct full SQL generation}
\begin{table}[H]
\centering
\scriptsize
\caption{Synthetic changed lines after canonical SQL formatting. The metric counts deleted plus inserted lines in a canonicalized diff between the original SQL and the generated SQL after patch application.}
\begin{tabular}{llrrrr}
\toprule
\textbf{Method} & \textbf{Output} & \textbf{Median} & \textbf{P75} & \textbf{P90} & \textbf{Max} \\
\midrule
Direct full SQL rewrite & Final SQL & 13 & 20 & 31 & 538 \\
Diff-ICL & Stage-1 patch & 12 & 18 & 29 & 92 \\
Diff-7B & Stage-1 patch & 16 & 23 & 32 & 89 \\
\bottomrule
\end{tabular}
\label{tab:synthetic-changed-lines}
\end{table}

\paragraph{Synthetic changed lines.}
We first compare rewrite scope at the line level after canonical SQL formatting. 
For direct full query generation, we canonicalize both the original SQL and the generated SQL, and then compute a synthetic diff between them. 
The metric counts the total number of deleted and inserted canonical SQL lines, and is therefore used only as a coarse measure of textual rewrite size.

Table~\ref{tab:synthetic-changed-lines} shows that patching does not uniformly shrink every rewrite. 
Direct full query generation and Diff-ICL have similar median changed-line counts, 13 and 12 respectively. 
The main difference appears in the tail: the maximum changed-line count reaches 538 for direct full query rewriting, compared with 92 for Diff-ICL and 89 for Diff-7B. 
This suggests that the patch representation primarily suppresses extreme broad rewrites, rather than forcing all optimizations to be small.

\label{app:patch vs end2end}
\lstset{
  basicstyle=\ttfamily\scriptsize,
  breaklines=true,
  columns=fullflexible,
  keepspaces=true,
  showstringspaces=false
}

\definecolor{boxgraybg}{RGB}{248,248,248}
\definecolor{boxgrayframe}{RGB}{210,210,210}


\section{Query Execution Plan Context}
\label{app:qep_context}

Query execution plans (QEPs) provide execution-level or optimizer-level evidence about how a database engine processes a SQL query. 
In our framework, QEPs are used as contextual information for identifying optimization opportunities. 

\paragraph{Dialect-specific plan extraction.}
Because QEP interfaces differ across database systems, we use the native plan extraction method supported by each dialect. 
For PostgreSQL, we use \texttt{EXPLAIN ANALYZE}. 
For MySQL, we use \texttt{EXPLAIN FORMAT=JSON}. 
For Microsoft SQL Server, we use \texttt{SET SHOWPLAN\_ALL ON}. 
For SQLite, we use \texttt{EXPLAIN QUERY PLAN}. 
For Oracle, we use \texttt{EXPLAIN PLAN FOR} followed by \texttt{DBMS\_XPLAN.DISPLAY}. 

\paragraph{Case study.}
Figures~\ref{fig:sql_case_study_appendix} and~\ref{fig:qep_rewrite_case} show a representative PostgreSQL example. 
The original query computes the final pricing field through a correlated scalar aggregation. 
The rewrite removes the redundant inner \texttt{orders} lookup and reformulates the computation as a set based aggregation over \texttt{lineitem}, followed by a \texttt{LEFT JOIN} back to \texttt{orders}. 
The QEP makes this optimization target explicit by exposing the repeated aggregate subplan in the original query.

\begin{figure}[t]
\centering

\begin{minipage}[t]{0.49\linewidth}
\begin{tcolorbox}[
  enhanced,
  colback=boxgraybg,
  colframe=boxgrayframe,
  boxrule=0.4pt,
  arc=2pt,
  left=3pt,right=3pt,top=3pt,bottom=3pt,
  title={Slow SQL},
  fonttitle=\bfseries,
  equal height group=sqlcase
]
\begin{lstlisting}[language=SQL,escapeinside={(*@}{@*)}]
SELECT
    TableActual.price AS "Final Pricing"
FROM
(
    SELECT
        o.o_orderkey,
        o.o_custkey,
        o.o_orderstatus,
        ...
        (
            SELECT
                COALESCE(
                    ROUND(SUM(l.l_extendedprice * (1 - l.l_discount)), 2),
                    0
                )
            FROM orders o2
            JOIN lineitem l
              ON o2.o_orderkey = l.l_orderkey
            WHERE o2.o_orderkey = o.o_orderkey
              AND l.l_linestatus = 'F'
        ) AS price,
        (*@\color{myred}\bfseries correlated scalar aggregation@*)
        ...
    FROM orders o
    GROUP BY
        o.o_orderkey,
        o.o_custkey,
        o.o_orderstatus
) AS TableActual
LEFT JOIN customer t1
  ON TableActual.customerkey = t1.c_custkey;
\end{lstlisting}
\end{tcolorbox}
\end{minipage}
\hfill
\begin{minipage}[t]{0.49\linewidth}
\begin{tcolorbox}[
  enhanced,
  colback=boxgraybg,
  colframe=boxgrayframe,
  boxrule=0.4pt,
  arc=2pt,
  left=3pt,right=3pt,top=3pt,bottom=3pt,
  title={Optimized SQL},
  fonttitle=\bfseries,
  equal height group=sqlcase
]
\begin{lstlisting}[language=SQL,escapeinside={(*@}{@*)}]
SELECT
    COALESCE(price_by_order.price, 0) 
        AS "Final Pricing"
FROM orders AS o
LEFT JOIN
(
    SELECT
        l.l_orderkey,
        COALESCE(
            ROUND(SUM(l.l_extendedprice * (1 - l.l_discount)), 2),
            0
        ) AS price
    FROM lineitem AS l
    WHERE l.l_linestatus = 'F'
    GROUP BY l.l_orderkey
) AS price_by_order
  ON price_by_order.l_orderkey = o.o_orderkey;

(*@\color{myred}\bfseries set based aggregation + LEFT JOIN@*)
\end{lstlisting}
\end{tcolorbox}
\end{minipage}

\caption{
SQL case study. 
The original query computes the final pricing field through a correlated scalar aggregation, while the rewritten query replaces it with set based aggregation over \texttt{lineitem} followed by a semantics-preserving \texttt{LEFT JOIN}.
}
\label{fig:sql_case_study_appendix}
\end{figure}

\begin{figure}[t]
\centering

\begin{minipage}[t]{0.49\linewidth}
\begin{tcolorbox}[
  enhanced,
  colback=boxgraybg,
  colframe=boxgrayframe,
  boxrule=0.4pt,
  arc=2pt,
  left=3pt,right=3pt,top=3pt,bottom=3pt,
  title={Original QEP},
  fonttitle=\bfseries,
  equal height group=qepcompare
]
\begin{lstlisting}[escapeinside={(*@}{@*)}]
Subquery Scan on tableactual
  -> Group
       Group Key: o.o_orderkey

       -> Index Scan on orders o

       SubPlan 1
       (*@\color{myred}\bfseries correlated subplan@*)

       -> Aggregate
            loops=4500000
            (*@\color{myred}\bfseries repeated aggregate@*)

            -> Nested Loop
                 loops=4500000
                 (*@\color{myred}\bfseries N+1 lookup pattern@*)

                 -> Index Only Scan on orders o2
                      Index Cond:
                        o_orderkey = o.o_orderkey

                 -> Index Scan on lineitem l
                      Index Cond:
                        l_orderkey = o.o_orderkey
                      Filter:
                        l_linestatus = 'F'
\end{lstlisting}
\end{tcolorbox}
\end{minipage}
\hfill
\begin{minipage}[t]{0.49\linewidth}
\begin{tcolorbox}[
  enhanced,
  colback=boxgraybg,
  colframe=boxgrayframe,
  boxrule=0.4pt,
  arc=2pt,
  left=3pt,right=3pt,top=3pt,bottom=3pt,
  title={Rewritten QEP},
  fonttitle=\bfseries,
  equal height group=qepcompare
]
\begin{lstlisting}[escapeinside={(*@}{@*)}]
Incremental Sort
  Sort Key:
    o.o_orderkey,
    o.o_custkey,
    o.o_orderstatus

  -> Merge Left Join
       Merge Cond:
         o.o_orderkey = l.l_orderkey

       -> Index Scan on orders o

       -> Finalize GroupAggregate
            Group Key: l.l_orderkey
            (*@\color{myred}\bfseries grouped aggregation@*)

            -> Gather Merge
                 -> Sort
                      Sort Key: l.l_orderkey

                      -> Partial HashAggregate
                           Group Key: l.l_orderkey
                           (*@\color{myred}\bfseries aggregate once by order@*)

                           -> Parallel Seq Scan on lineitem l
                                Filter:
                                  l_linestatus = 'F'
\end{lstlisting}
\end{tcolorbox}
\end{minipage}

\caption{
QEP case study for correlated aggregation rewriting. 
The original QEP exposes a correlated scalar aggregate subplan executed once per \texttt{orders} row, forming an N+1 lookup pattern. 
The rewritten QEP removes this correlated subplan and replaces it with a set based aggregation over \texttt{lineitem}, grouped by \texttt{l\_orderkey}, followed by a \texttt{LEFT JOIN} back to \texttt{orders}.
}
\label{fig:qep_rewrite_case}
\end{figure}

\section{Evaluation metrics}
\label{app:preliminaries_formal}

We evaluate each optimized query $q'$ along two metrics: \textbf{$\Delta$EX(Execution Accuracy Degradation)} and \textbf{R-VES(Reward-based Valid Efficiency Score)}. Let $Q = \{q_i\}_{i=1}^{N}$ be the evaluation set, $m_i \in \{0,1\}$ the match indicator (whether $q_i'$ executes successfully and produces equivalent results to $q_i$ on database instance $D$), and $\tau_i = {T(q_i; D)}/{T(q_i'; D)}$ the speedup ratio.

\paragraph{EX.} EX measures the fraction of optimized queries that are semantically equivalent to the original queries. Non-executable rewrites are assigned zero match.
\begin{equation}
    \mathrm{EX} = \frac{1}{N}\sum_{i=1}^{N} m_i
\end{equation}

\paragraph{$\Delta$EX.}
$\Delta$EX measures the degradation in execution accuracy:
\begin{equation}
    \Delta\mathrm{EX} = 1 - \mathrm{EX} .
\end{equation}

\paragraph{R-VES.} R-VES measures the average efficiency gain score over all queries, where the score is determined by both semantic correctness and the degree of speedup. It extends the VES metric~\citep{li2023can} to ensure greater stability and reliability. The formulation is:
\begin{equation}
    \mathrm{R\text{-}VES} = \frac{1}{N}\sum_{i=1}^{N} m_i \cdot g(\tau_i)
\end{equation}
where
\begin{equation}
    g(\tau)=
    \begin{cases}
    1.00, & \tau \ge 2,\\
    0.75, & 1.2 \le \tau < 2,\\
    0.50, & 1.0 \le \tau < 1.2,\\
    0.25, & 0.5 \le \tau < 1.0,\\
    0, & \tau < 0.5.
    \end{cases}
\end{equation}

Compared to the original R-VES formulation, our R-VES introduces two modifications. First, we remove the $1.25$ bonus tier for $\tau \ge 2$ and cap the maximum score at $1.0$, so that R-VES is bounded in $[0,1]$. Second, we introduce $1.2\times$ as a new tier boundary that separates the original $[1, 2)$ interval into $[1.0, 1.2)$ and $[1.2, 2.0)$. This reflects the empirical observation that speedups below $1.2\times$ cannot be reliably distinguished from execution-time noise under practical evaluation protocols (see Section~\ref{app:speedup_threshold}). Queries in this noise-dominated regime receive a reduced score of $0.50$ rather than the full $0.75$, preserving partial evaluation signal while avoiding conflation of measurement variance with genuine optimization.

\subsection{Justification of the 1.2$\times$ speedup threshold}
\label{app:speedup_threshold}

We use a 1.2$\times$ speedup threshold in two places: first, when filtering Slow-Fast SQL pairs during data construction, and second, when assigning efficiency credit in R-VES. 
The purpose of this threshold is to separate reliable efficiency gains from execution-time fluctuations. 
We empirically justify the 1.2$\times$ boundary by quantifying the execution-time variance inherent in repeated SQL evaluation. 
We select 40 TPC-H queries from the benchmark and execute each query 20 times across four scale factors (SF = 0.1, 1, 3, 10) on the same database instance, yielding 160 measurement groups in total. 
For each group of 20 runs, we compute all pairwise ratios between runs of the same query to measure the false speedup that arises purely from execution noise.

\begin{table}[H]
\centering
\caption{
Probability that execution noise alone produces a false speedup exceeding different thresholds, estimated from pairwise ratios across 20 repeated runs of 40 SQL queries.
}
\label{tab:false_speedup_threshold}
\small
\begin{tabular}{cc}
\toprule
\textbf{Threshold} & \textbf{P(false speedup $\geq$ threshold)} \\
\midrule
1.05$\times$ & 10.63\% \\
1.10$\times$ & 5.15\% \\
1.15$\times$ & 3.33\% \\
1.20$\times$ & 2.49\% \\
1.25$\times$ & 1.97\% \\
\bottomrule
\end{tabular}
\end{table}

As shown in Table~\ref{tab:false_speedup_threshold}, the probability that pure execution noise produces a false speedup $\geq$ 1.2$\times$ is approximately 2.5\%, placing the 1.2$\times$ threshold near the 97.5th percentile of the noise distribution ($\alpha=0.025$). 
This supports using 1.2$\times$ as a conservative boundary for reliable speedup. 
In data construction, this threshold helps ensure that retained Slow-Fast pairs reflect genuine optimization rather than measurement variance. 
In R-VES, it prevents marginal speedups in the noise-dominated range from being treated as fully reliable efficiency gains. 
Queries with speedup in $[1.0, 1.2)$ therefore receive reduced credit, preserving partial evaluation signal without conflating execution noise with true optimization.

\subsection{Speedup ratio}
\label{app:speedup-ratio}

To quantify the gap between model rewrites and the reference
expert rewrites in Effi-SQL, we adopt the speedup ratio (SR) metric
from SWE-fficiency~\citep{ma2026swefficiency}. If the rewrite is executable and semantically matched, the instance speedup ratio is computed as
\[
\mathrm{SR}_i =
\frac{\mathrm{Speedup}^{sys}_i}{\mathrm{Speedup}^{expert}_i},
\]
where
\[
\mathrm{Speedup}^{expert}_i = \frac{T_i^{orig}}{T_i^{expert}},
\qquad
\mathrm{Speedup}^{sys}_i = \frac{T_i^{orig}}{T_i^{sys}}.
\]
For failed rewrites, including non-executable outputs and semantic mismatches, we assign
\[
\mathrm{SR}_i = \frac{1}{\mathrm{Speedup}^{expert}_i}.
\]

Following SWE-fficiency, we aggregate instance-level SR values
using the harmonic mean:
\[
\mathrm{SR}
=
\frac{N}
{\sum_{i=1}^{N} \frac{1}{\mathrm{SR}_i}}.
\]
The harmonic mean is appropriate for averaging ratios and penalizes
systems that achieve high speedups on only a small subset of instances
while performing poorly on others. Since it is particularly sensitive
to extreme lower-tail values, we apply one-sided $1\%$ lower-tail
winsorization before aggregation:
\[
\widetilde{\mathrm{SR}}_i =
\max(\mathrm{SR}_i, Q_{0.01}),
\qquad
\widetilde{\mathrm{SR}}
=
\frac{N}
{\sum_{i=1}^{N} \frac{1}{\widetilde{\mathrm{SR}}_i}},
\]
where \(Q_{0.01}\) denotes the first percentile of instance-level SR values.

As shown in Table~\ref{tab:sr_hm_results}, we report results only for
the direct full SQL generation setting. Current frontier models still
exhibit a substantial efficiency gap relative to the reference expert
rewrites, leaving considerable room for further improvement.
\begin{table}[H]
\centering
\caption{
\textbf{Speedup Ratio (SR)} and execution outcomes on the PostgreSQL subset of Effi-SQL.
SR is defined as the model speedup normalized by the reference expert
speedup. For failed rewrites, we assign the score
$1/\mathrm{Speedup}_{\mathrm{expert}}$.
We report the value after one-sided 1\% lower-tail winsorization.
\textbf{Higher SR} is better. 
}\label{tab:sr_hm_results}
\scriptsize
\setlength{\tabcolsep}{6pt}
\renewcommand{\arraystretch}{1.1}
\begin{tabular}{lcccc}
\toprule
\multirow{2}{*}{\textbf{System}} &
\multirow{2}{*}{\textbf{SR ($\times$) $\uparrow$}} &
\multicolumn{3}{c}{\textbf{Passes Correctness Tests}} \\
\cmidrule(lr){3-5}
&
&
\textbf{Slower than Original (\%)$\downarrow$} &
\textbf{Faster than Original (\%)$\uparrow$} &
\textbf{Faster than Expert (\%)$\uparrow$} \\
\midrule
Claude-Opus-4.5          & \textbf{0.111575}$\times$ & 12.67 & \underline{38.00} & \textbf{43.33} \\
DeepSeek-V3              & \underline{0.061636}$\times$ & 12.00 & 34.00 & 36.67 \\
Claude-Opus-4.6          & 0.053569$\times$ & 14.00 & \textbf{39.67} & 37.00 \\
GLM-4.7                  & 0.049080$\times$ & 10.33 & 31.00 & \underline{37.67} \\
Kimi-K2-Thinking         & 0.020844$\times$ &  \textbf{5.00} & 19.67 & 33.33 \\
MiniMax-M2               & 0.020425$\times$ &  \underline{8.67} & 26.00 & 28.67 \\
Qwen3-Coder              & 0.019304$\times$ & 10.00 & 31.67 & 31.67 \\
Qwen-2.5-7B-Instruct     & 0.005051$\times$ & 18.00 & 24.67 & 13.67 \\
\bottomrule
\end{tabular}
\end{table}
\section{TPC-H scale factor selection}
\label{app:scale_stability}
\subsection{Metric definition}

We select the evaluation scale by balancing two goals: preserving the optimization landscape observed at SF10 and keeping evaluation cost practical. To this end, we consider four types of evidence: optimization gain, fidelity to SF10, execution cost, and repeated-run stability.

For each query pair $i$ at scale $s$, let $T^{\mathrm{slow}}_{i,s}$ and $T^{\mathrm{opt}}_{i,s}$ denote the execution times of the original and optimized queries, respectively. We define raw speedup as
\begin{equation}
r_{i,s}=\frac{T^{\mathrm{slow}}_{i,s}}{T^{\mathrm{opt}}_{i,s}}.
\end{equation}
Because speedup is ratio-valued and often heavy-tailed, we summarize optimization gain primarily with the median speedup and, when needed, the geometric mean speedup. In this analysis, speedup statistics are descriptive rather than decisive: they indicate how much optimization benefit is observed at each scale, but they do not by themselves determine which scale is the best proxy for SF10.

To assess whether a smaller scale preserves the optimization landscape of SF10, we use two fidelity measures. First, we compute the Spearman rank correlation between query-level speedups at scale $s$ and those at SF10:
\begin{equation}
\rho_s=\mathrm{corr}\big(\mathrm{rank}(r_{1,s},\dots,r_{N,s}),\mathrm{rank}(r_{1,10},\dots,r_{N,10})\big).
\end{equation}
A higher $\rho_s$ indicates that scale $s$ better preserves the relative ordering of which queries benefit more or less from optimization under SF10. Second, we measure per-query deviation from SF10 using the absolute log-ratio error
\begin{equation}
d_{i,s}=\left|\log\frac{r_{i,s}}{r_{i,10}}\right|.
\end{equation}
Smaller $d_{i,s}$ means that the query-level speedup at scale $s$ is closer to the corresponding behavior observed at SF10.

To quantify evaluation cost, we define the total execution time of a query pair as
\begin{equation}
C_{i,s}=T^{\mathrm{slow}}_{i,s}+T^{\mathrm{opt}}_{i,s}.
\end{equation}
We summarize the typical successful-query cost using the median execution time. To reflect practical evaluation burden under the 5-minute timeout limit, we additionally report timeout-aware expected attempt cost over all attempted query pairs:
\begin{equation}
\bar C_s=\frac{1}{M}\sum_{i=1}^{M} C^{\mathrm{att}}_{i,s},
\end{equation}
where
\begin{equation}
C^{\mathrm{att}}_{i,s}=
\begin{cases}
T^{\mathrm{slow}}_{i,s}+T^{\mathrm{opt}}_{i,s}, & \text{if the query pair succeeds and matches},\\
300\text{s}, & \text{if a timeout occurs}.
\end{cases}
\end{equation}
This metric captures not only successful execution time but also the additional burden induced by timeouts at larger scales.

Finally, we evaluate repeated-run stability using the coefficient of variation (CV) of speedup across repeated executions. Lower CV indicates that the observed optimization gain is less sensitive to runtime noise. We report the median speedup CV across the repeated-run subset as a summary of typical stability.

\subsection{Results}

Across 131 candidate query pairs, only 100 were successfully executed and result-matched across all four scales. We therefore use this common 100-query subset for all paired cross-scale comparisons involving speedup, fidelity, and successful-query execution time. The remaining roughly 30 failed cases are excluded from paired fidelity analysis but are still used to characterize practical evaluation burden, including timeout rate and timeout-aware expected attempt cost. This separation avoids selection bias in fidelity comparisons while preserving information about real evaluation difficulty.

Table~\ref{tab:core_scale_results} summarizes the matched-subset comparison. The main selection criterion is not which scale maximizes every raw performance statistic, but which smaller scale most faithfully reproduces the SF10 optimization landscape at acceptable cost. Under this criterion, SF3 is the strongest candidate. Its rank fidelity to SF10 is clearly the highest among smaller scales, with Spearman correlation 0.9472, compared with 0.8711 for SF1 and 0.8374 for SF0.1. The same conclusion is supported by the median absolute log-ratio deviation, which is smallest for SF3 at 0.0715, versus 0.1256 for SF1 and 0.1229 for SF0.1. At the query level, SF3 is closer to SF10 than SF1 on 58 of the 100 shared queries, and closer than SF0.1 on 66 of them. These results indicate that SF3 best preserves both the ordering and magnitude of query-level optimization gains relative to SF10.

Raw speedup statistics are consistent with this conclusion but should be interpreted as secondary evidence. SF0.1 attains the highest median speedup, while SF3 remains competitive at 1.4763 and exceeds both SF1 and SF10. This suggests that SF3 does not sacrifice optimization gain in exchange for fidelity. More importantly, SF3 retains this fidelity at lower cost than SF10: its median successful-query execution time is 20.42s, compared with 60.08s for SF10, or roughly one-third of the SF10 cost. SF1 is cheaper still, but its fidelity to SF10 is noticeably weaker.

Table~\ref{tab:stability_results} reports repeated-run stability together with practical burden over all 131 attempted query pairs. Here, SF0.1 and SF3 emerge as the two strongest practical candidates, but for different reasons. SF0.1 has the lowest speedup variability, lowest timeout rate, and lowest expected attempt cost, making it highly attractive for cheap exploratory runs. However, its much weaker rank fidelity and larger log-ratio deviation show that it does not preserve the SF10 optimization landscape as reliably as SF3. SF3, by contrast, combines near-best stability with stronger fidelity to SF10, while still remaining far less costly and much less failure-prone than directly evaluating at SF10. SF10 exhibits the largest practical burden, with an 18.32\% timeout rate and an expected attempt cost of 174.46s.

Overall, SF0.1 and SF3 represent two different operating points: SF0.1 optimizes for cost and stability, whereas SF3 optimizes for fidelity to SF10 while remaining practical. Since our goal is not merely to minimize runtime but to choose a realistic benchmark evaluation scale that preserves high-scale optimization behavior, SF3 provides the best fidelity--cost trade-off. We therefore use SF3 as the main evaluation scale, while treating SF0.1 as a low-cost exploratory setting rather than a faithful high-scale proxy.

\begin{table}[t]
\centering
\caption{Matched-subset fidelity and speedup comparison on the 100 query pairs that succeed and match across all four scales. Bold indicates values favorable to SF3 as a practical SF10 proxy.}
\label{tab:core_scale_results}
\small
\setlength{\tabcolsep}{6pt}
\renewcommand{\arraystretch}{1.12}
\begin{tabular}{lcccc}
\toprule[1pt]
Scale & Median Speedup & Spearman w/ SF10 & Median $d_{i,s}$ & Median Exec. Time (s) \\
\midrule[0.6pt]
SF0.1 & \textbf{1.5778} & 0.8374 & 0.1229 & \textbf{0.67} \\
SF1   & 1.3883 & 0.8711 & 0.1256 & 7.19 \\
SF3   & 1.4763 & \textbf{0.9472} & \textbf{0.0715} & 20.42 \\
SF10  & 1.4324 & --- & --- & 60.08 \\
\bottomrule[1pt]
\end{tabular}
\end{table}

\begin{table}[t]
\centering
\caption{All-attempt stability and practical evaluation cost. Speedup CV is computed from repeated executions and summarizes runtime-induced speedup variability. Timeout rate and expected attempt cost are computed over all 131 attempted query pairs. Bold indicates values favorable to SF3 as a practical evaluation scale.}
\label{tab:stability_results}
\small
\setlength{\tabcolsep}{7pt}
\renewcommand{\arraystretch}{1.12}
\begin{tabular}{lccc}
\toprule[1pt]
Scale & Median Speedup CV & Timeout Rate & Expected Attempt Cost (s) \\
\midrule[0.6pt]
SF0.1 & \textbf{0.0265} & \textbf{1.53\%} & \textbf{10.00} \\
SF1   & 0.0360 & 5.34\% & 37.64 \\
SF3   & \textbf{0.0275} & 6.11\% & 77.91 \\
SF10  & 0.0378 & 18.32\% & 174.46 \\
\bottomrule[1pt]
\end{tabular}
\end{table}

\section{Reward formulation}
\label{app:reward}
\paragraph{Execution-grounded reward for the Constraint Aligner.}
We train the Constraint Aligner with an execution-grounded reward that focuses on patch validity, executability, and semantic preservation. 
Given an original query $q$ and a model-generated patch $\hat{p}$, we first extract a unified diff patch from the model output and apply it to the original SQL query to obtain the patched query $\hat{q}=\mathrm{Apply}(q,\hat{p})$. 
If the response does not contain an extractable patch, the patch contains no effective edit, or the patch cannot be applied to the original query, the output receives a reward of $-1.0$.

We additionally detect no-op or revert behavior by normalizing the original and patched SQL, removing comments, whitespace differences, trailing semicolons, and case distinctions. 
If the patched SQL is equivalent to the original SQL under this normalization, the output receives a small negative reward of $-0.1$. 
This case is treated separately because it is usually syntactically valid and semantics-preserving, but it does not perform the optimization task: the model either makes no substantive edit or reverts the query to the original form. 
Assigning a positive reward to such cases would encourage a degenerate conservative policy in which the aligner avoids semantic risk by undoing optimization patches. 
At the same time, we do not penalize no-op or revert outputs as harshly as malformed patches or execution errors, since they are safer than invalid or semantically incorrect rewrites.

For nontrivial patched queries, we execute $\hat{q}$ against the target PostgreSQL database under a bounded statement timeout. 
If execution raises a SQL error, the reward is $-1.0$. 
If execution times out, the reward is $+0.1$, reflecting that the query is syntactically valid but not useful as an efficient repair. 
For successful executions, we compare the result set against the precomputed result hash of the original query. 
If the result hash matches and the patch is not a no-op, the output receives the maximum reward of $+1.0$. 
If execution succeeds but the result set differs, the output receives $-0.5$. 
Database connection failures or internal reward-code failures receive a neutral reward of $0.0$ and are logged for inspection.

Formally, the reward is:
\begin{equation}
R(q, \tilde{p}) = \begin{cases}
+1.0, & \text{semantic match} \\
+0.1, & \text{execution timeout} \\
+0.0, & \text{system error} \\
-0.1, & \text{no-op or revert} \\
-0.5, & \text{semantic mismatch} \\
-1.0, & \text{malformed patch or execution error}
\end{cases}
\end{equation}
This reward is intentionally constraint-oriented. 
It does not directly optimize latency; instead, it trains the aligner to produce executable and semantics-preserving patches while discouraging malformed patches, empty edits, no-op repairs, and semantic drift.

\section{Multi-dialect evaluation of Diff-RL}

SQL dialects differ in syntax, built-in functions, and execution semantics, so a rewrite valid in one engine may fail in another.
To test whether Diff-RL generalizes across dialects, we apply the same framework to MySQL and SQL Server, both using Qwen-2.5-7B-Instruct as the base model.
As shown in Table~\ref{tab:multidialect}, Diff-RL consistently improves both $\Delta$EX and R-VES across dialects, suggesting that the framework is not specific to PostgreSQL.

\begin{table}[H]
\centering
\caption{Multi-dialect results.}
\label{tab:multidialect}
\scriptsize
\renewcommand{\arraystretch}{1.03}
\setlength{\tabcolsep}{5pt}
\begin{tabular}{lll l}
\toprule
\textbf{Dialect} & \textbf{Metric} & \textbf{Base} & \textbf{Diff-RL} \\
\midrule
\multirow{2}{*}{MySQL}
& $\Delta$EX $\downarrow$ & 64.00 & \textbf{54.00}\loss{-10.0} \\
& R-VES $\uparrow$ & 19.88 & \textbf{30.75}\gain{+10.87} \\
\bottomrule
\end{tabular}
\end{table}

\section{Error Analysis Details}
\label{app:error_analysis_details}

This appendix keeps one representative example for each error category and shows the corresponding base SQL, failed rewrite or patch, corrected Diff-ICL output when applicable, and failure analysis.

\subsection{Example 1: Direct Full-SQL Execution Error, Diff-ICL Correct}

{\small
\begin{longtable}{P{0.19\linewidth}P{0.77\linewidth}}
\toprule
Field & Content \\
\midrule
\endfirsthead
\toprule
Field & Content \\
\midrule
\endhead
\field{Failure Mode} & End-to-end returns compare\_failed: \texttt{Exec Error: column "buckets" does not exist}. Diff-ICL step 2 is ok with speedup 2.4927. \\
\field{Source SQL} & \sqlblock{
\sqlline{SELECT '<=30' AS buckets FROM lineitem WHERE l_quantity <= 30 GROUP BY buckets}
\sqlline{UNION ALL SELECT '31-40' AS buckets FROM lineitem WHERE l_quantity > 30 AND l_quantity <= 40 GROUP BY buckets}
\sqlline{UNION ALL SELECT '41-50' AS buckets FROM lineitem WHERE l_quantity > 40 AND l_quantity <= 50 GROUP BY buckets}
\sqlline{UNION ALL SELECT '>50' AS buckets FROM lineitem WHERE l_quantity > 50 GROUP BY buckets}
} \\
\field{Broken Full SQL} & \sqlblock{
\sqlline{SELECT}
\sqlline{  CASE}
\sqlline{    WHEN l_quantity <= 30 THEN '<=30'}
\sqlline{    WHEN l_quantity <= 40 THEN '31-40'}
\sqlline{    WHEN l_quantity <= 50 THEN '41-50'}
\sqlline{    ELSE '>50'}
\sqlline{  END AS buckets}
\sqlline{FROM lineitem}
\sqlline{GROUP BY CASE WHEN l_quantity <= 30 THEN '<=30' WHEN l_quantity <= 40 THEN '31-40' WHEN l_quantity <= 50 THEN '41-50' ELSE '>50' END}
\sqlredline{ORDER BY CASE buckets}
\sqlline{  WHEN '<=30' THEN 1 WHEN '31-40' THEN 2}
\sqlline{  WHEN '41-50' THEN 3 WHEN '>50' THEN 4 END;}
} \\
\field{Diff-ICL Patch} & \sqlblock{
\sqlline{--- original.sql}
\sqlline{+++ optimized.sql}
\sqlline{@@ -1 +1,14 @@}
\sqlline{-SELECT '<=30' AS buckets FROM lineitem WHERE l_quantity <= 30 GROUP BY buckets UNION ALL SELECT '31-40' AS buckets FROM lineitem WHERE l_quantity > 30 AND l_quantity <= 40 GROUP BY buckets UNION ALL SELECT '41-50' AS buckets FROM lineitem WHERE l_quantity > 40 AND l_quantity <= 50 GROUP BY buckets UNION ALL SELECT '>50' AS buckets FROM lineitem WHERE l_quantity > 50 GROUP BY buckets}
\sqlcolorline{+SELECT buckets}
\sqlcolorline{+FROM (}
\sqlline{+  SELECT CASE}
\sqlline{+    WHEN l_quantity <= 30 THEN '<=30'}
\sqlline{+    WHEN l_quantity <= 40 THEN '31-40'}
\sqlline{+    WHEN l_quantity <= 50 THEN '41-50'}
\sqlline{+    ELSE '>50'}
\sqlline{+  END AS buckets}
\sqlline{+  FROM lineitem}
\sqlline{+) t}
\sqlcolorline{+GROUP BY buckets}
\sqlcolorline{+ORDER BY CASE buckets}
\sqlline{+  WHEN '<=30' THEN 1 WHEN '31-40' THEN 2}
\sqlline{+  WHEN '41-50' THEN 3 WHEN '>50' THEN 4 END}
} \\
\field{Why This Fits} & The full-SQL rewrite tries to order by the SELECT alias inside a CASE expression that PostgreSQL cannot resolve. Step 2 refines the structure by materializing the bucket alias in a subquery, then grouping and ordering on the real derived column. \\
\bottomrule
\end{longtable}
}

\clearpage

\subsection{Example 2: Direct Full-SQL Semantic Mismatch, Diff-ICL Correct}

{\small
\begin{longtable}{P{0.19\linewidth}P{0.77\linewidth}}
\toprule
Field & Content \\
\midrule
\endfirsthead
\toprule
Field & Content \\
\midrule
\endhead
\field{Failure Mode} & End-to-end is labeled pred\_mismatch\_content with the same row count. Diff-ICL step 2 is ok with speedup 2.23. \\
\field{Source SQL} & \sqlblock{
\sqlline{WITH ordered_users AS MATERIALIZED (}
\sqlline{  SELECT c.c_name, c.c_address, EXTRACT(YEAR FROM AGE(CURRENT_DATE, CAST('1970-01-01' AS DATE))) AS age, c.c_custkey}
\sqlline{  FROM customer c}
\sqlline{)}
\sqlline{SELECT c_name, c_address, age FROM ordered_users ORDER BY c_custkey ASC;}
} \\
\field{Mismatching Full SQL} & \sqlblock{
\sqlline{SELECT}
\sqlline{  c_name,}
\sqlline{  c_address,}
\sqlredline{  (SELECT EXTRACT(YEAR FROM AGE(}
\sqlredline{     CURRENT_DATE,}
\sqlredline{     DATE '1970-01-01'}
\sqlredline{  )))::INTEGER AS age}
\sqlline{FROM customer}
\sqlline{ORDER BY c_custkey;}
} \\
\field{Diff-ICL Patch} & \sqlblock{
\sqlline{--- original.sql}
\sqlline{+++ optimized.sql}
\sqlline{@@ -1 +1,6 @@}
\sqlline{-WITH ordered_users AS MATERIALIZED (SELECT c.c_name, c.c_address, EXTRACT(YEAR FROM AGE(CURRENT_DATE, CAST('1970-01-01' AS DATE))) AS age, c.c_custkey FROM customer c) SELECT c_name, c_address, age FROM ordered_users ORDER BY c_custkey ASC;}
\sqlcolorline{+SELECT}
\sqlline{+  c.c_name,}
\sqlline{+  c.c_address,}
\sqlline{+  EXTRACT(YEAR FROM AGE(CURRENT_DATE, CAST('1970-01-01' AS DATE))) AS age}
\sqlcolorline{+FROM customer c}
\sqlcolorline{+ORDER BY c.c_custkey ASC;}
} \\
\field{Why This Fits} & The direct rewrite changes the age expression into a scalar subquery plus cast, which is enough for content mismatch. Step 2 performs the intended minimal rewrite: remove CTE materialization, keep the original expression, and preserve ordering. \\
\bottomrule
\end{longtable}
}

\clearpage

\subsection{Example 3: Diff-ICL Step 1 (Patch Generator) Syntax Error, Step 2 (Constraint Aligner) Fixed}

{\small
\begin{longtable}{P{0.19\linewidth}P{0.77\linewidth}}
\toprule
Field & Content \\
\midrule
\endfirsthead
\toprule
Field & Content \\
\midrule
\endhead
\field{Failure Mode} & Step 1 is compare\_failed: \texttt{Exec Error: syntax error at or near "ORDER"}. Step 2 is ok with speedup 4.0018. \\
\field{Source SQL} & \sqlblock{
\sqlline{WITH cutoff_date AS ( SELECT MIN(o_orderdate) AS min_date FROM orders )}
\sqlline{SELECT r.o_orderkey, r.o_custkey, r.o_orderstatus, r.o_totalprice,}
\sqlline{       r.o_orderdate, r.o_orderpriority, r.o_clerk, r.o_shippriority, r.o_comment}
\sqlline{FROM (}
\sqlline{  SELECT orders.*, ROW_NUMBER() OVER(ORDER BY o_orderdate DESC, O_ORDERKEY DESC) as rn}
\sqlline{  FROM orders JOIN cutoff_date ON orders.o_orderdate = cutoff_date.min_date}
\sqlline{) AS r}
\sqlline{WHERE rn <= 3}
\sqlline{ORDER BY o_orderdate ASC, O_ORDERKEY ASC}
} \\
\field{Step 1 Patch} & \sqlblock{
\sqlline{--- original.sql}
\sqlline{+++ optimized.sql}
\sqlline{@@ -1,18 +1,14 @@}
\sqlline{-WITH cutoff_date AS ( SELECT MIN(o_orderdate) AS min_date FROM orders )}
\sqlline{ SELECT}
\sqlline{-  r.o_orderkey, r.o_custkey, r.o_orderstatus, r.o_totalprice,}
\sqlline{-  r.o_orderdate, r.o_orderpriority, r.o_clerk, r.o_shippriority, r.o_comment}
\sqlline{-FROM (}
\sqlline{-  SELECT orders.*, ROW_NUMBER() OVER(ORDER BY o_orderdate DESC, O_ORDERKEY DESC) as rn}
\sqlline{-  FROM orders JOIN cutoff_date ON orders.o_orderdate = cutoff_date.min_date}
\sqlline{-) AS r}
\sqlline{-WHERE rn <= 3}
\sqlline{-ORDER BY o_orderdate ASC, O_ORDERKEY ASC}
\sqlline{+  o_orderkey, o_custkey, o_orderstatus, o_totalprice,}
\sqlline{+  o_orderdate, o_orderpriority, o_clerk, o_shippriority, o_comment}
\sqlline{+FROM orders}
\sqlline{+WHERE o_orderdate = (SELECT MIN(o_orderdate) FROM orders)}
\sqlredline{+ORDER BY o_orderdate DESC, o_orderkey DESC}
\sqlredline{+LIMIT 3}
\sqlline{+ORDER BY o_orderdate ASC, o_orderkey ASC;}
} \\
\field{Step 2 Patch} & \sqlblock{
\sqlline{--- original.sql}
\sqlline{+++ optimized.sql}
\sqlline{@@ -1,20 +1,19 @@}
\sqlline{-WITH cutoff_date AS ( SELECT MIN(o_orderdate) AS min_date FROM orders )}
\sqlline{ SELECT}
\sqlline{-  r.o_orderkey, r.o_custkey, r.o_orderstatus, r.o_totalprice,}
\sqlline{-  r.o_orderdate, r.o_orderpriority, r.o_clerk, r.o_shippriority, r.o_comment}
\sqlline{+  o_orderkey, o_custkey, o_orderstatus, o_totalprice,}
\sqlline{+  o_orderdate, o_orderpriority, o_clerk, o_shippriority, o_comment}
\sqlline{ FROM (}
\sqlline{   SELECT}
\sqlline{-    orders.*,}
\sqlline{-    ROW_NUMBER() OVER(ORDER BY o_orderdate DESC, O_ORDERKEY DESC) as rn}
\sqlline{-  FROM orders JOIN cutoff_date ON orders.o_orderdate = cutoff_date.min_date}
\sqlcolorline{-) AS r}
\sqlline{-WHERE rn <= 3}
\sqlline{+    *}
\sqlcolorline{+  FROM orders}
\sqlline{+  WHERE o_orderdate = (SELECT MIN(o_orderdate) FROM orders)}
\sqlcolorline{+  ORDER BY O_ORDERKEY DESC}
\sqlline{+  LIMIT 3}
\sqlline{+) AS r}
\sqlline{ ORDER BY o_orderdate ASC, O_ORDERKEY ASC}
} \\
\field{Why This Fits} & Step 1 places a second ORDER BY after LIMIT, which is invalid SQL. Step 2 wraps the top-3 selection in a subquery and applies the final presentation order outside, so the refinement is both syntactic and semantic. \\
\bottomrule
\end{longtable}
}

\clearpage

\subsection{Example 4: Diff-ICL Step 2 (Constraint Aligner) Refinement Failure}

{\small
\begin{longtable}{P{0.19\linewidth}P{0.77\linewidth}}
\toprule
Field & Content \\
\midrule
\endfirsthead
\toprule
Field & Content \\
\midrule
\endhead
\field{Failure Mode} & Step 1 is ok; step 2 becomes compare\_failed: \texttt{Exec Error: syntax error at or near "dt"}. \\
\field{Base SQL} & \sqlblock{
\sqlline{SELECT date_trunc('day', dt_col)::date AS dt, to_char(SUM(1) / 60.0, 'FM999999999.00') as Hours}
\sqlline{FROM generate_series('2015-01-27 07:32:35.000'::timestamp, '2015-04-29 14:39:35.000'::timestamp, '1 minute'::interval) AS dt_col}
\sqlline{GROUP BY date_trunc('day', dt_col)}
\sqlline{ORDER BY date_trunc('day', dt_col);}
} \\
\field{Step 1 Patch} & \sqlblock{
\sqlline{--- original.sql}
\sqlline{+++ optimized.sql}
\sqlline{@@ -1 +1,7 @@}
\sqlline{-SELECT date_trunc('day', dt_col)::date AS dt, to_char(SUM(1) / 60.0, 'FM999999999.00') as Hours FROM generate_series('2015-01-27 07:32:35.000'::timestamp, '2015-04-29 14:39:35.000'::timestamp, '1 minute'::interval) AS dt_col GROUP BY date_trunc('day', dt_col) ORDER BY date_trunc('day', dt_col);}
\sqlcolorline{+SELECT dt, to_char(cnt / 60.0, 'FM999999999.00') as Hours}
\sqlline{+FROM (}
\sqlcolorline{+  SELECT date_trunc('day', dt_col)::date AS dt, SUM(1) as cnt}
\sqlline{+  FROM generate_series('2015-01-27 07:32:35.000'::timestamp, '2015-04-29 14:39:35.000'::timestamp, '1 minute'::interval) AS dt_col}
\sqlline{+  GROUP BY 1}
\sqlcolorline{+) sub}
\sqlline{+ORDER BY dt;}
} \\
\field{Step 2 Patch} & \sqlblock{
\sqlline{--- original.sql}
\sqlline{+++ optimized.sql}
\sqlline{@@ -1 +1,20 @@}
\sqlline{-SELECT date_trunc('day', dt_col)::date AS dt, to_char(SUM(1) / 60.0, 'FM999999999.00') as Hours FROM generate_series('2015-01-27 07:32:35.000'::timestamp, '2015-04-29 14:39:35.000'::timestamp, '1 minute'::interval) AS dt_col GROUP BY date_trunc('day', dt_col) ORDER BY date_trunc('day', dt_col);}
\sqlline{+WITH date_range AS (}
\sqlline{+  SELECT generate_series('2015-01-27'::date, '2015-04-29'::date, '1 day')::date AS d,}
\sqlline{+         '2015-01-27 07:32:35'::timestamp AS start_ts,}
\sqlline{+         '2015-04-29 14:39:35'::timestamp AS end_ts}
\sqlline{+), daily_minutes AS (}
\sqlline{+  SELECT d::date AS dt, CASE}
\sqlline{+    WHEN d = start_ts::date AND d = end_ts::date THEN EXTRACT(EPOCH FROM (end_ts - start_ts)) / 60}
\sqlline{+    WHEN d = start_ts::date THEN EXTRACT(EPOCH FROM ((d + '1 day'::interval) - start_ts)) / 60}
\sqlline{+    WHEN d = end_ts::date THEN EXTRACT(EPOCH FROM (end_ts - d::timestamp)) / 60}
\sqlline{+    ELSE 1440}
\sqlline{+  END AS mins}
\sqlline{+  FROM date_range}
\sqlline{+)}
\sqlredline{+SELECT dt,}
\sqlredline{+       to_char(mins / 60.0,}
\sqlredline{+               'FM999999999.00') AS Hours}
\sqlredline{+FROM daily_minutes}
\sqlredline{+ORDER BY dt;}
} \\
\field{Why This Fits} & The first patch is a conservative equivalent rewrite. The second-stage refinement attempts a more ambitious date-range computation and fails at execution, so this is a clean example of refinement making a passing patch worse. \\
\bottomrule
\end{longtable}
}

\clearpage

\subsection{Example 5: 7B Patch-Only / Diff-RL Remaining Failure}

{\small
\begin{longtable}{P{0.19\linewidth}P{0.77\linewidth}}
\toprule
Field & Content \\
\midrule
\endfirsthead
\toprule
Field & Content \\
\midrule
\endhead
\field{Failure Mode} & 7B Diff-RL result is compare\_failed: \texttt{Exec Error: missing FROM-clause entry for table "os"}. \\
\field{Source SQL} & \sqlblock{
\sqlline{WITH order_scores AS (}
\sqlline{  SELECT o.o_orderkey AS order_id, o.o_orderpriority, o.o_orderdate, l.l_suppkey, l.l_extendedprice}
\sqlline{  FROM orders o JOIN lineitem l ON o.o_orderkey = l.l_orderkey}
\sqlline{  WHERE o.o_orderkey IN (SELECT l_orderkey FROM lineitem WHERE l_suppkey = 2)}
\sqlline{)}
\sqlline{SELECT row_number() over() AS num,}
\sqlline{       (array_agg(s.s_name ORDER BY os.l_extendedprice DESC) FILTER (WHERE os.l_suppkey != 2))[1] AS opponent_name,}
\sqlline{       MAX(CASE WHEN os.l_suppkey != 2 THEN os.l_extendedprice END) AS opponent_score,}
\sqlline{       MAX(CASE WHEN os.l_suppkey = 2 THEN os.l_extendedprice END) AS player_score}
\sqlline{FROM order_scores os JOIN supplier s ON os.l_suppkey = s.s_suppkey}
\sqlline{GROUP BY os.order_id, os.o_orderdate::date}
\sqlline{ORDER BY os.order_id ASC;}
} \\
\field{7B Remaining Failure} & \sqlblock{
\sqlline{### Analysis:}
\sqlline{The flawed patch incorrectly removed the CTE and attempted to push the filter condition into a LATERAL join,}
\sqlredline{but failed to properly reference the CTE alias 'os' in the main query.}
\sqlline{### Verified Patch:}
\sqlline{--- original.sql}
\sqlline{+++ optimized.sql}
\sqlline{@@ -1,25 +1,17 @@}
\sqlline{-WITH order_scores AS ( ... )}
\sqlline{+SELECT row_number() over() AS num, ...}
\sqlredline{+-- remaining query still refers to os.l_extendedprice and os.l_suppkey}
} \\
\field{Why This Fits} & The model recognizes that the patch has alias and CTE problems, but the generated patch trajectory still leaves references to os after restructuring the query. This is a representative remaining failure after small-model refinement. \\
\bottomrule
\end{longtable}
}

\clearpage

\subsection{Example 6: Model Hesitation and Wrong Final SQL}

This subsection focuses on model-side output instability. The response hesitates between analysis, schema-level suggestions, and SQL rewrites; instead of cleanly committing to the correct rewrite, it eventually emits an executable SQL block that is wrong.

{\small
\begin{longtable}{P{0.19\linewidth}P{0.77\linewidth}}
\toprule
Field & Content \\
\midrule
\endfirsthead
\toprule
Field & Content \\
\midrule
\endhead
\field{Failure Mode} & The model response is unstable: it alternates between reasoning, SQL, and schema/index suggestions, then ultimately emits a wrong executable SQL answer. The evaluated run fails with \texttt{Exec Error: syntax error at or near "set"}. \\
\field{Base SQL} & \sqlblock{
\sqlline{SELECT DISTINCT inspectmode as Category,}
\sqlline{       FIRST_VALUE(inspectres) OVER (PARTITION BY inspectmode ORDER BY inspectres) as Details}
\sqlline{FROM inspection;}
} \\
\field{Model Behavior} & The important point is that the model never cleanly converges to a single final SQL answer; its output trajectory remains hesitant and contaminated by non-answer material before ending on an incorrect SQL answer. \\
\field{Final Wrong SQL} & \sqlblock{
\sqlline{SELECT inspectmode as Category,}
\sqlline{       FIRST_VALUE(inspectres) OVER (PARTITION BY inspectmode ORDER BY inspectres) as Details}
\sqlline{FROM inspection}
\sqlredline{ORDER BY inspectres;}
} \\
\field{Why Wrong} & The correct simplification is to project \texttt{inspectmode} and \texttt{inspectres} directly. The final answer keeps window-function-style ordering behavior and changes the result contract, so it is not a faithful rewrite of the source query. \\
\field{Takeaway} & This example is kept separate from ordinary SQL execution errors because the failure is driven by model-side response instability: the model produces a hesitant trajectory and ends on a wrong SQL block rather than a clean optimized answer. \\
\bottomrule
\end{longtable}
\subsection{Example 7: SQL Extraction Failure}

This subsection covers the missing extraction-failure case. The model does not provide the requested optimized SQL rewrite; it only emits a schema-level DDL suggestion, so there is no valid rewritten query answer to evaluate.

{\small
\begin{longtable}{P{0.19\linewidth}P{0.77\linewidth}}
\toprule
Field & Content \\
\midrule
\endfirsthead
\toprule
Field & Content \\
\midrule
\endhead
\field{Failure Mode} & The run is labeled \texttt{sql\_extract\_failed}. The model output contains reasoning and a DDL statement, but no valid optimized SELECT rewrite. \\
\field{Source SQL} & \sqlblock{
\sqlline{SELECT sitekey AS id, sitelabel AS forename, sitelabel AS surname}
\sqlline{FROM (}
\sqlline{  SELECT sitekey, sitelabel, COUNT(*) OVER (PARTITION BY sitelabel) AS name_count}
\sqlline{  FROM plants}
\sqlline{) t}
\sqlline{WHERE name_count > 1;}
} \\
\field{Model Output} & \sqlblock{
\sqlline{<think>}
\sqlline{The plan shows a sequential scan and sort for the window partition.}
\sqlline{The model proposes an access-path optimization rather than a SQL rewrite.}
\sqlline{</think>}
\sqlline{```sql}
\sqlredline{CREATE INDEX idx_plants_sitelabel ON plants(sitelabel);}
\sqlline{```}
} \\
\field{Why This Is Extract Fail} & The task asks for an optimized SQL query. The model instead outputs \texttt{CREATE INDEX idx\_plants\_sitelabel ON plants(sitelabel);}. This may be a performance suggestion, but it is not a replacement query and therefore leaves no valid full-SQL rewrite to extract. \\
\field{Takeaway} & This case isolates extraction failure from semantic mismatch and execution failure: the model response never supplies the required answer artifact. \\
\bottomrule
\end{longtable}

\clearpage
\section{Prompt}
\label{app:prompt}
\begin{figure}[H]
\centering
\includegraphics[draft=false,width=1.0\linewidth,height=0.95\textheight,keepaspectratio]{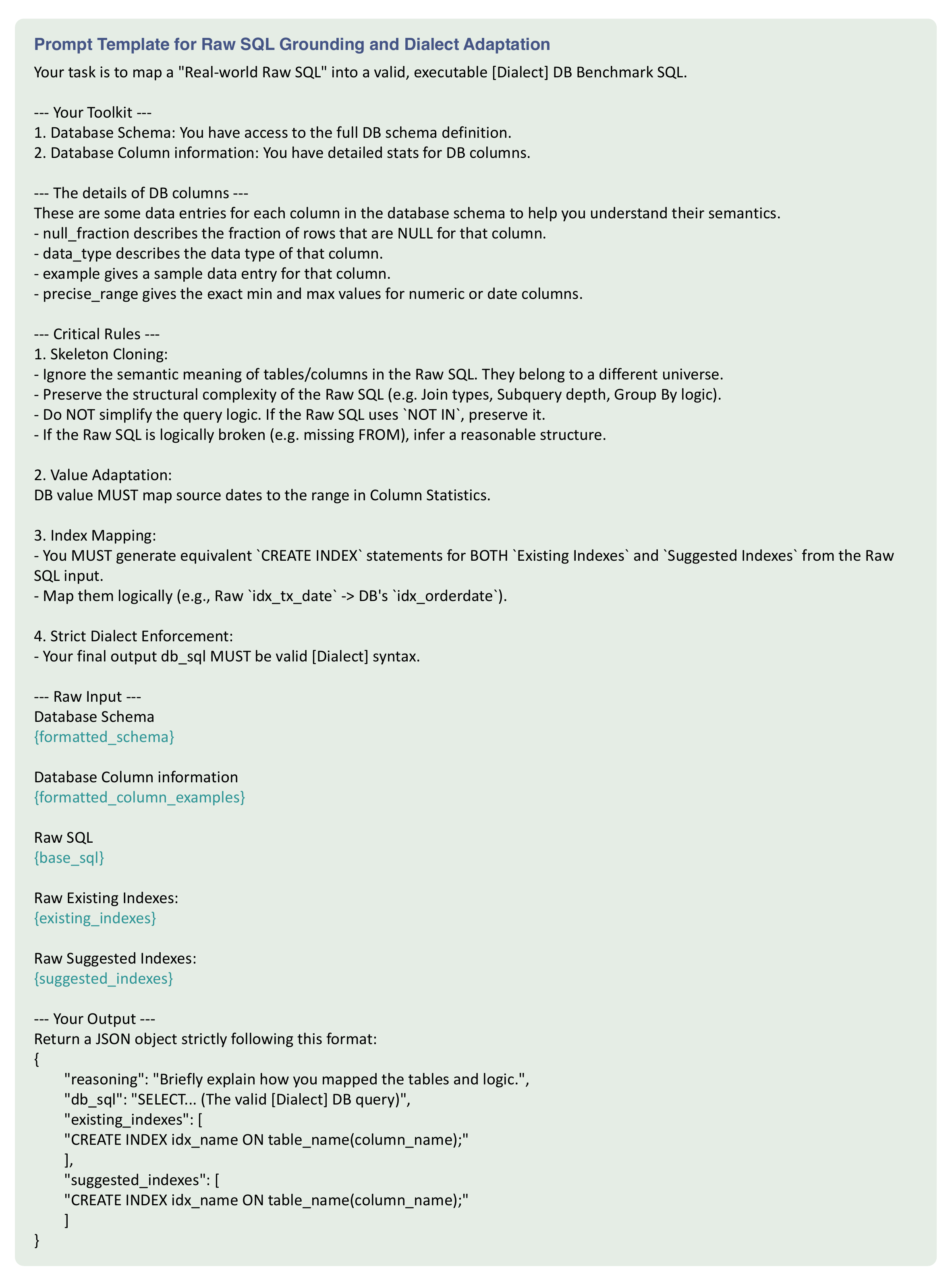}
\caption{Prompt Template for Raw SQL Grounding and Dialect Adaptation.}
\label{fig:raw-sql-grounding}
\end{figure}

\begin{figure}[p]
\centering
\includegraphics[draft=false,width=1.0\linewidth,height=0.95\textheight,keepaspectratio]{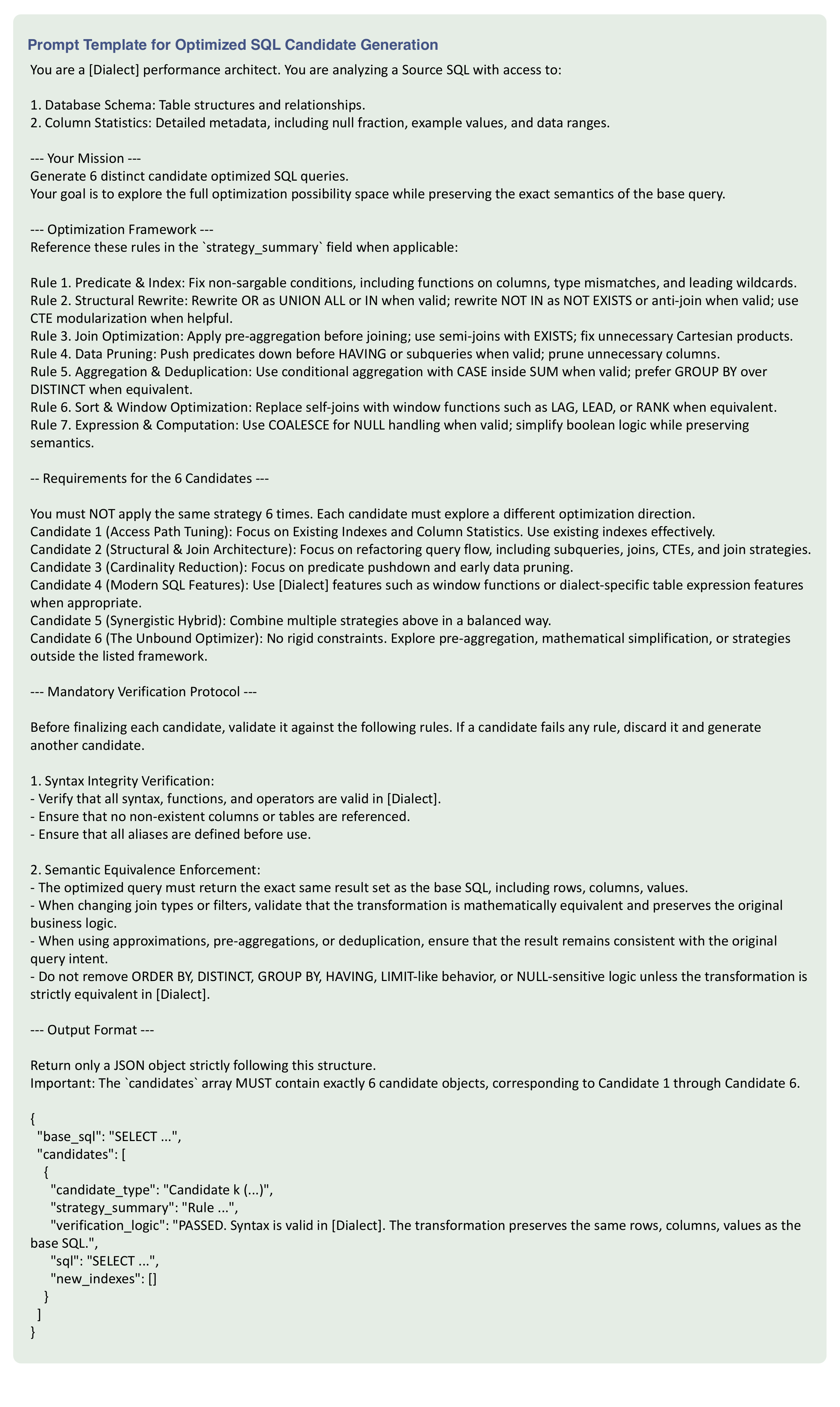}
\caption{Prompt template for optimized SQL candidate generation.}
\label{fig:candidate-generate}
\end{figure}

\begin{figure}[p]
\centering
\includegraphics[draft=false,width=1.0\linewidth,height=0.95\textheight,keepaspectratio]{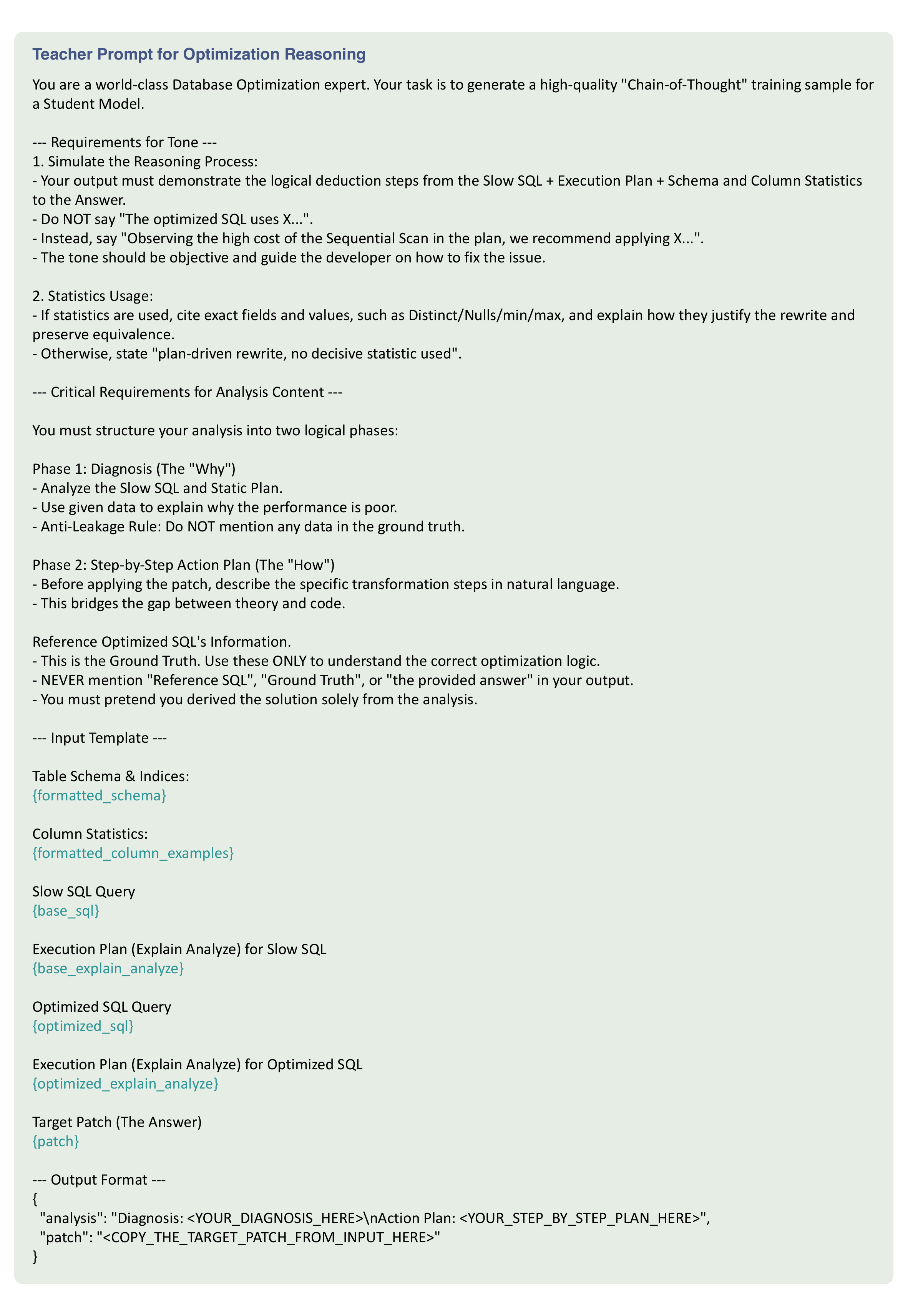}
\caption{Teacher Prompt for Optimization Reasoning.}
\label{fig:optimization-reasoning-generation}
\end{figure}

\begin{figure}[p]
\centering
\includegraphics[draft=false,width=1.0\linewidth,height=0.95\textheight,keepaspectratio]{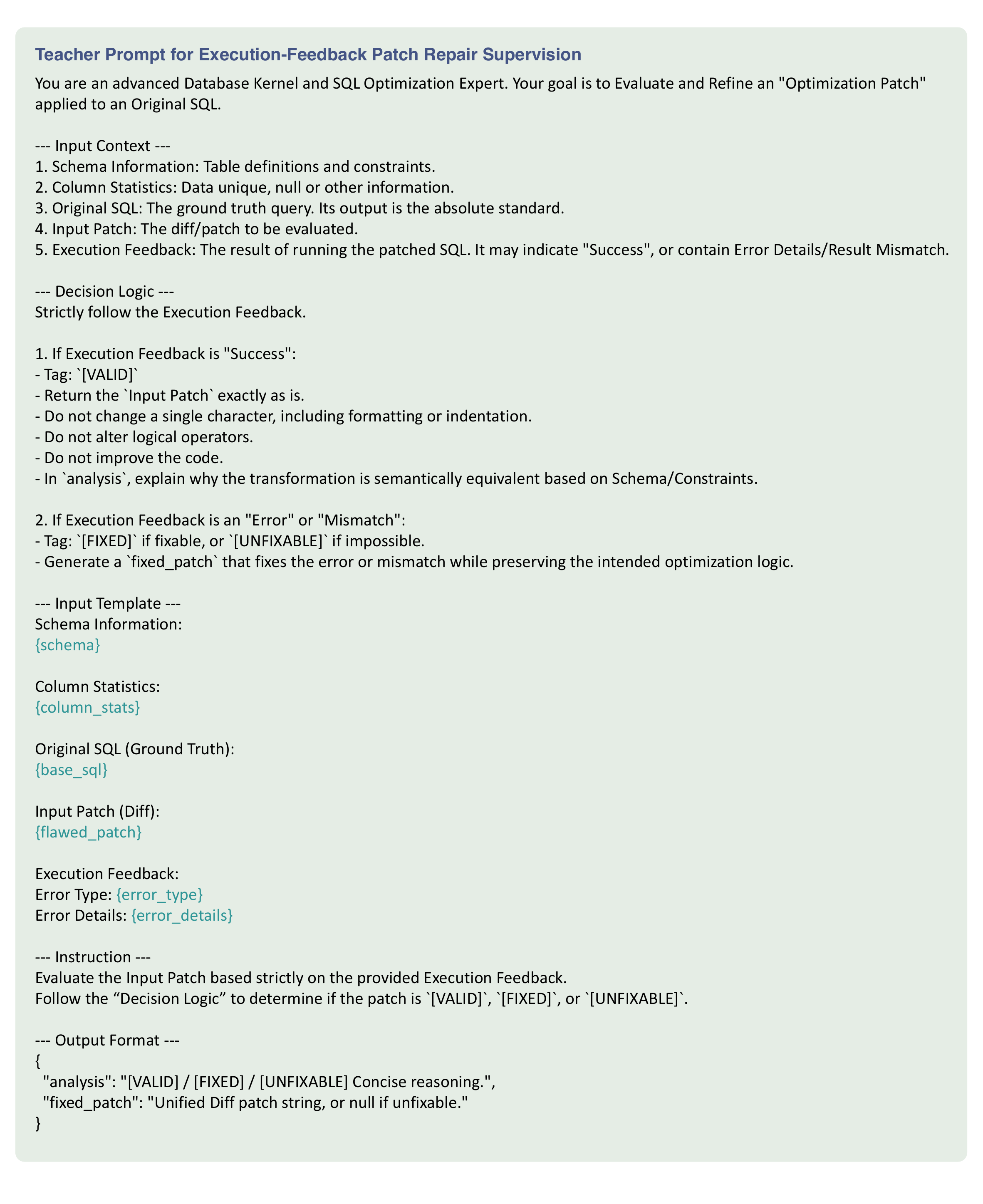}
\caption{Teacher Prompt for Execution-Feedback Patch Repair Supervision.}
\label{fig:fix-analysis-generation}
\end{figure}

\begin{figure}[p]
\centering
\includegraphics[draft=false,width=1.0\linewidth,height=0.95\textheight,keepaspectratio]{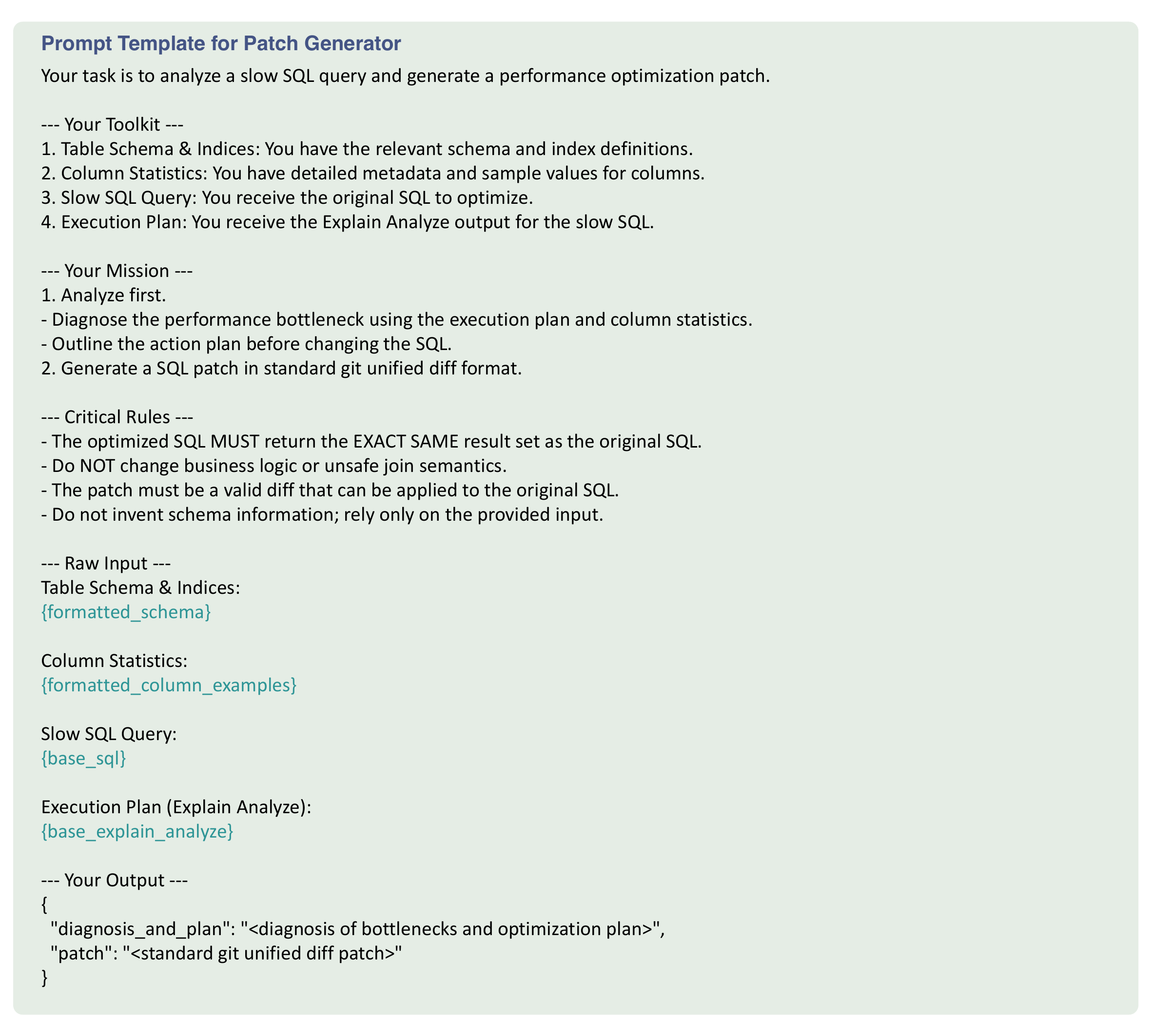}
\caption{Prompt Template for Patch Generator.}
\label{fig:patch-generator}
\end{figure}

\begin{figure}[p]
\centering
\includegraphics[draft=false,width=1.0\linewidth,height=0.95\textheight,keepaspectratio]{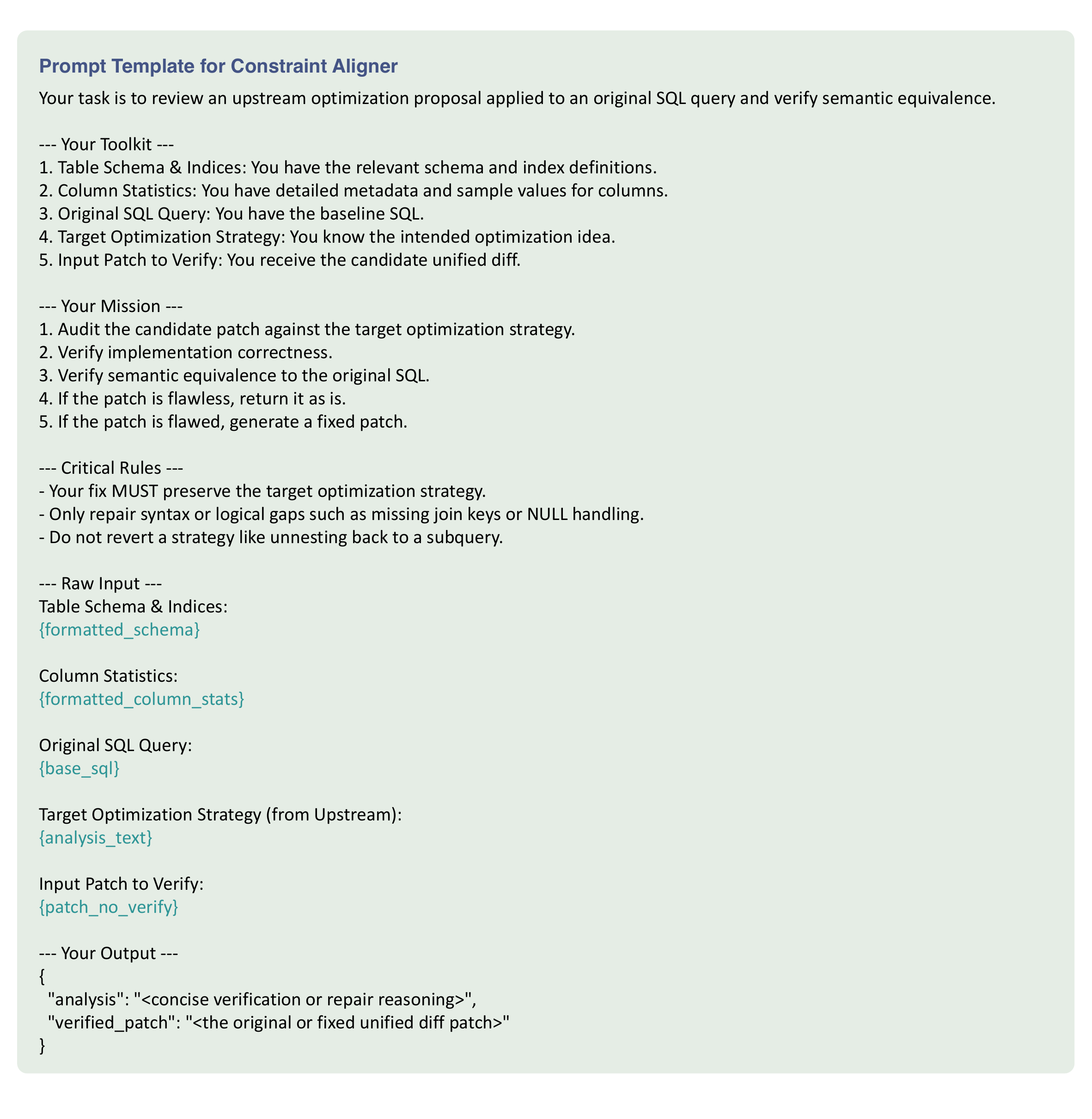}
\caption{Prompt Template for Constraint Aligner.}
\label{fig:constraint-aligner}
\end{figure}

\begin{figure}[p]
\centering
\includegraphics[draft=false,width=1.0\linewidth,height=0.95\textheight,keepaspectratio]{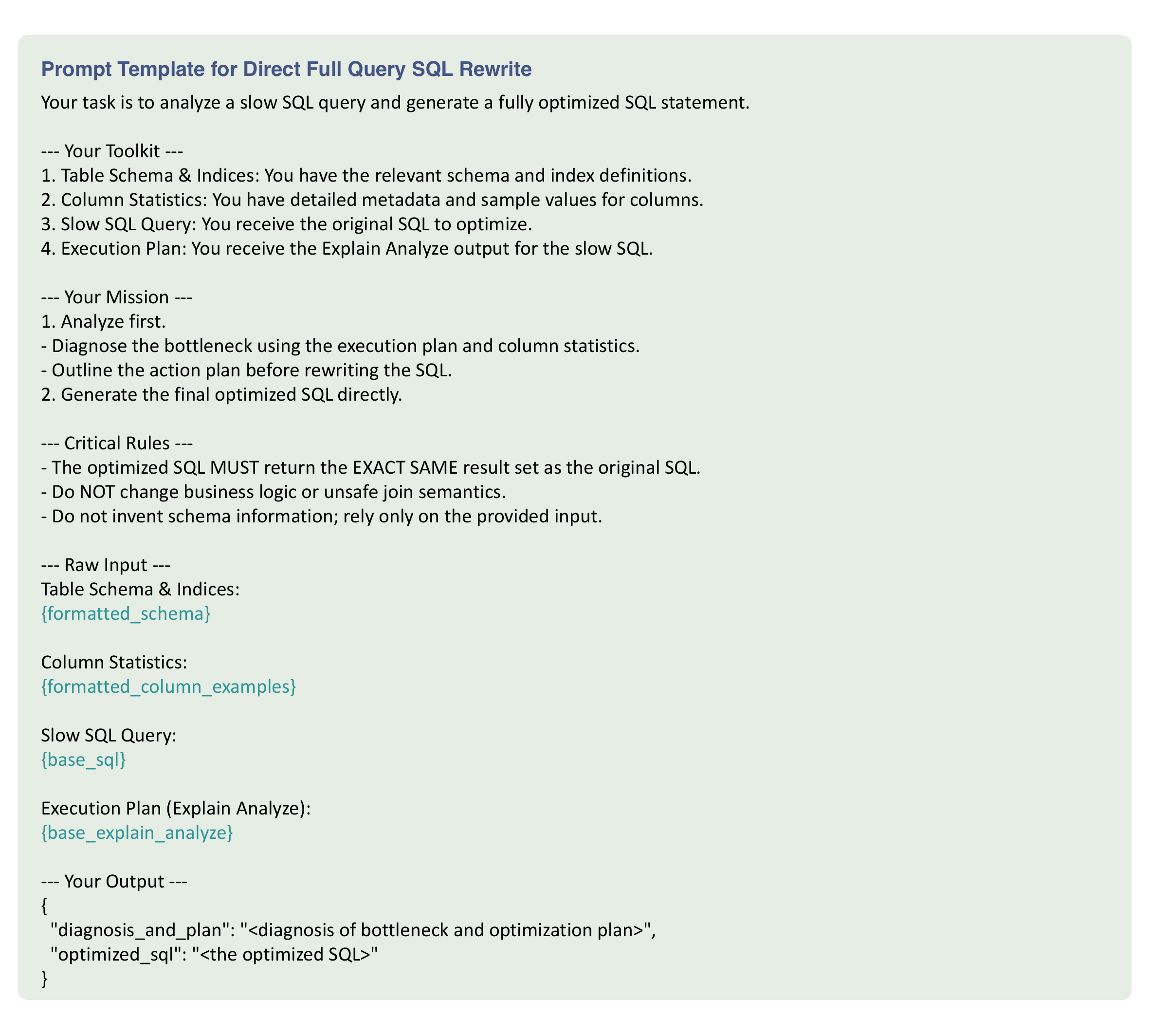}
\caption{Prompt Template for Direct Full Query SQL Rewrite.}
\label{fig:end2end}
\end{figure}



\end{document}